\documentclass[iop,numberedappendix]{emulateapj}

\usepackage[export]{adjustbox}
\usepackage{times}
\usepackage{commath}
\usepackage[utf8x]{inputenc}
\usepackage{hyperref}
\hypersetup{
  pdftitle={Impact of atmospheric chromatic effects on weak lensing measurements},
  pdfauthor={Joshua Meyers},
  colorlinks=true,
  linkcolor=blue,
  citecolor=blue,
  urlcolor=blue,
  bookmarksnumbered
}

\newcommand{\red}[1]{\textcolor{red}{#1}}

\newcommand{\happy}{\text{\,\includegraphics[height=1.25em,raise=-0.4em]{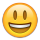}\,}}
\newcommand{\bored}{\text{\,\includegraphics[height=1.25em,raise=-0.4em]{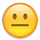}\,}}
\newcommand{\sad}{\text{\,\includegraphics[height=1.25em,raise=-0.4em]{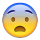}\,}}
\newcommand{\angry}{\text{\,\includegraphics[height=1.25em,raise=-0.4em]{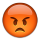}\,}}

\let\Re\undefined
\DeclareMathOperator{\Re}{Re}
\let\Im\undefined
\DeclareMathOperator{\Im}{Im}
\DeclareMathOperator{\Var}{Var}
\DeclareMathOperator{\Covar}{Covar}

\shorttitle{Weak-lensing atmospheric chromatic effects}
\shortauthors{Meyers et al.}

\begin{document}

\title{Impact of atmospheric chromatic effects on weak lensing measurements}

\author{Joshua E.~Meyers and Patricia R.~Burchat}
\affil{Kavli Institute for Particle Astrophysics and Cosmology, Department of Physics, Stanford University, Stanford, CA 94305}
\email{jmeyers314@gmail.com}

\begin{abstract}
Current and future imaging surveys will measure cosmic shear with statistical precision that demands a deeper understanding of potential systematic biases in galaxy shape measurements than has been achieved to date.
We use analytic and computational techniques to study the impact on shape measurements of two atmospheric chromatic effects for ground-based surveys such as the Dark Energy Survey and the Large Synoptic Survey Telescope (LSST): (i) atmospheric differential chromatic refraction and (ii) wavelength dependence of seeing.
We investigate the effects of using the point spread function (PSF) measured with stars to determine the shapes of galaxies that have different spectral energy distributions than the stars.
We find that both chromatic effects lead to significant biases in galaxy shape measurements for current and future surveys, if not corrected.
Using simulated galaxy images, we find a form of chromatic `model bias' that arises when fitting a galaxy image with a model that has been convolved with a stellar, instead of galactic, point spread function.
We show that both forms of atmospheric chromatic biases can be predicted (and corrected) with minimal model bias by applying an ordered set of perturbative PSF-level corrections based on machine-learning techniques applied to six-band photometry.
Catalog-level corrections do not address the model bias.
We conclude that achieving the ultimate precision for weak lensing from current and future ground-based imaging surveys requires a detailed understanding of the wavelength dependence of the PSF from the atmosphere, and from other sources such as optics and sensors.
The source code for this analysis is available at \url{https://github.com/DarkEnergyScienceCollaboration/chroma}.
\end{abstract}


\section{Introduction} %1
\label{sec:intro}
One of the principal goals of large astronomical imaging surveys is to constrain cosmological parameters by measuring the small departure from statistical isotropy of the shapes and orientations of distant galaxies, induced by the gravitational lensing from foreground large-scale structure. 
The observed shapes of galaxies, however, are not only affected by cosmic shear (typically a $\lesssim 1\%$ shift in the major-to-minor axis ratio), but are also determined by the combined point spread function (PSF) due to the atmosphere (for ground-based instruments), telescope optics, and the image sensor -- together often contributing a few $\%$ shift.
The size and shape of this additional convolution kernel is typically determined from the observed images of stars, which are effectively point sources before being smeared by the PSF. 
Galaxy images can then be deconvolved with the estimated convolution kernel, or alternatively, statistics derived from the convolved galaxy image can be corrected for the estimated convolution kernel.
Implicit in this approach is the assumption that the kernel for galaxies is the same as the kernel for stars.
If the PSF is dependent on wavelength, this assumption is violated since stars and galaxies have different spectral energy distributions (SEDs) and hence different PSFs.

In this paper, we consider two wavelength-dependent contributions to the PSF due to the atmosphere.

\begin{enumerate} 
  \item %1
  Differential chromatic refraction (DCR) --
  As photons enter and pass through the Earth's atmosphere, the refractive index changes from precisely 1 in vacuum to slightly more than 1 at the telescope. 
  This change leads to a small amount of refraction\footnote{We will use the terms `refraction' and `refraction angle' to mean the change in zenith angle due to refraction.  Note that this is distinct from `angle of refraction,' which is commonly defined with respect to the normal of the refracting interface.}, which depends on the zenith angle and the wavelength of the incoming photon. 
  For the range of zenith angles planned for imaging surveys, the effect is about 1~arcsecond of refraction for every degree away from zenith.

  For monochromatic sources, this change in the photon angle from above to below the atmosphere induces a very small zenith-direction flattening of images -- a few parts in $10^4$.
  Fortunately, stars and galaxies will be equally distorted, and in fact, this 0th-order effect will be removed entirely when fitting a World Coordinate System to the image using the known positions of reference stars.
  For sources with a range of wavelengths, however, we must consider the dispersive nature of atmospheric refraction -- referred to as differential chromatic refraction.
  Bluer photons are refracted slightly more than redder photons, as illustrated with the solid colored curves in Figure \ref{fig:DCRfilters}, where the relative amount of refraction is plotted as a function of wavelength for different zenith angles.
  We also show, as an illustrative example, the total transmission function for an airmass of 1.2 (including atmosphere, reflective and refractive optics, and CCD quantum efficiency) for the six filters planned for the Large Synoptic Survey Telescope (LSST)\footnote{LSST filter throughputs are available at \url{https://dev.lsstcorp.org/cgit/LSST/sims/throughputs.git/tree/baseline}}.
   LSST will primarily rely on $r$- and $i$-band for cosmic shear measurements.
  We see that at a zenith angle of 35 degrees, which is approximately the median expected zenith angle for the main `wide-fast-deep' part of the LSST survey \citep[Fig.\,3.3]{LSSTSB}, photons with wavelengths at opposite ends of the $r$-band filter will, on average, be separated on the focal plane by about 0.2 arcseconds or a third of the full-width-at-half-maximum (FWHM) of the typical PSF.
  The effect is smaller in $i$-band, with photons on opposite ends of the filter landing about 0.09 arcseconds apart when the zenith angle is 35 degrees.
  
  Figure \ref{fig:photon_landings} shows the effect of DCR on the PSF.  
  The left panels show the SED for a G5V star (top) and an Sa galaxy\footnote{We choose a G5V stellar SED, which is relatively blue, and an Sa galaxy SED, which is relatively red, to illustrate a case in which the biases are relatively large, but still representative.} redshifted to $z=0.6$ (bottom), and the wavelength distributions of surviving photons for each of these SEDs after being attenuated by the atmosphere, (LSST) filters, optics, and sensors.
  The right panels show the distribution of refraction angles for these same surviving photons.
  This distribution becomes an additional convolution kernel for the final PSF along the zenith axis, which illustrates 
that DCR will affect the shape of the PSF and the shape will depend on the spectral energy distribution (SED) of the star or galaxy.
  To compare the DCR kernel to the seeing kernel, a Moffat profile\footnote{A circular Moffat profile has  functional form $I(r) \propto (1+(r/\alpha)^2)^{-\beta}$, where $\alpha$ sets the profile size, and $\beta$ is a parameter that adjusts the importance of the profile core relative to the profile wings.  
  In the limiting case $\beta \rightarrow \infty$, the Moffat profile becomes a Gaussian profile.} with FWHM of $0.7$ arcseconds, which is typical of the expected seeing for LSST $r$- and $i$-band \citep[Fig.\,3.3]{LSSTSB}, is also plotted.
  
  In addition to changing the shape of the PSF and hence the shape of the PSF-convolved-galaxy image, DCR leads to a shift in the centroid of the PSF that depends on both the zenith angle {\em and} the object's SED. 
  If uncorrected, these centroid shifts can lead to problems registering exposures taken at different zenith and parallactic angles, which may manifest as additional blurring of the galaxy profile beyond what is accounted for by the individual epoch PSFs.

  \begin{figure*}
    \begin{center}
      \epsscale{1.0}
      \plotone{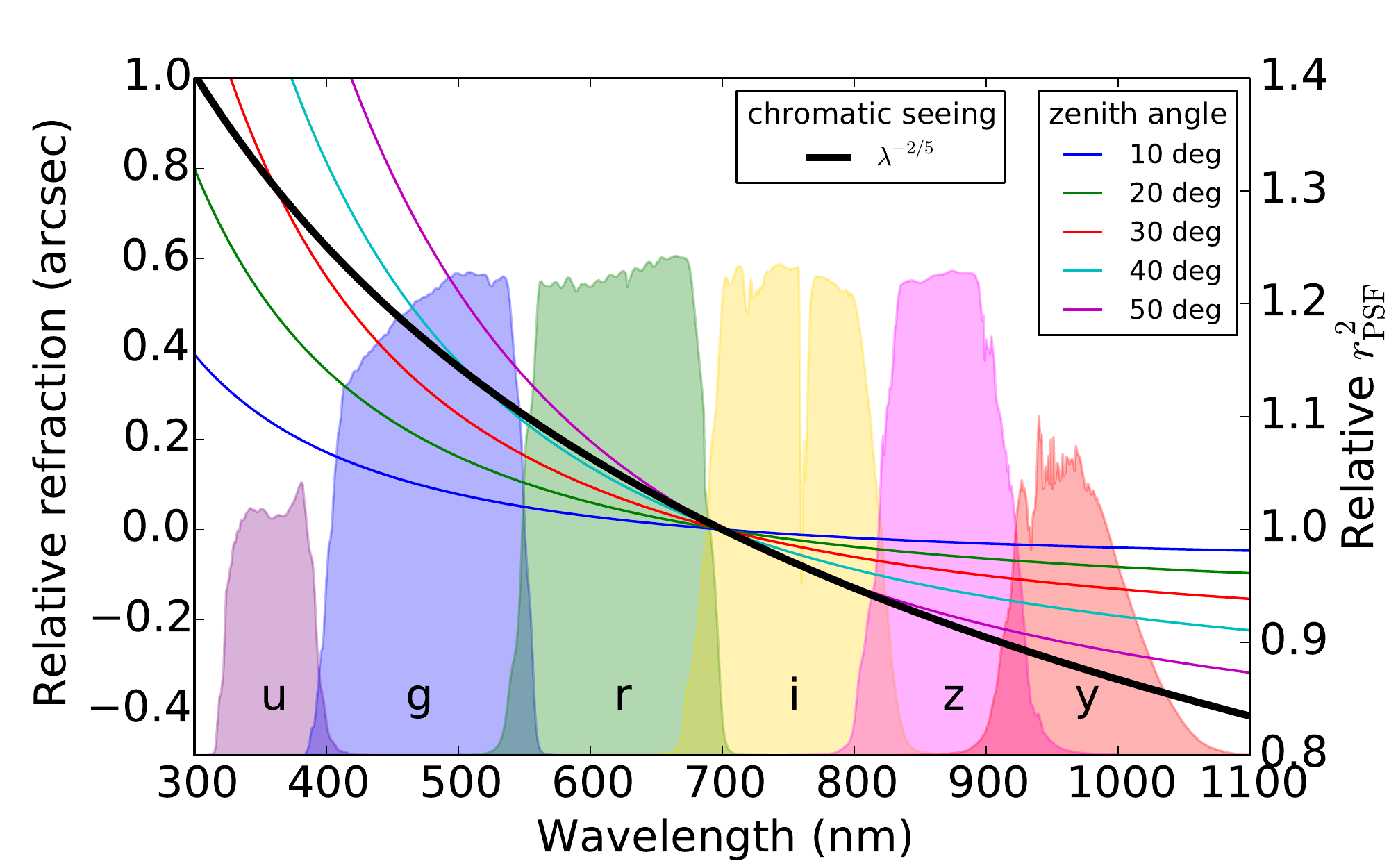}
    \end{center}
    \caption{
      \label{fig:DCRfilters} 
      Relative amount of differential chromatic refraction (left scale; solid colored curves) as a function of wavelength for different zenith angles, and the expected $\lambda^{-2/5}$ dependence (right scale; solid black curve) of the second-moment squared radius $r^2_{\rm PSF}$ for atmospheric seeing, defined in Eq.~\ref{eqn:r2}. 
      For each zenith angle, the refraction is set to an arbitrary value of 0 at a wavelength of 700~nm; $r^2_{\rm PSF}$ is measured relative to that at 700~nm.
      The total response function at an airmass of 1.2 for each of the six LSST passbands is overlaid in color.
    }
  \end{figure*}

  \begin{figure*}
    \begin{center}
      \epsscale{1.0}
      \plotone{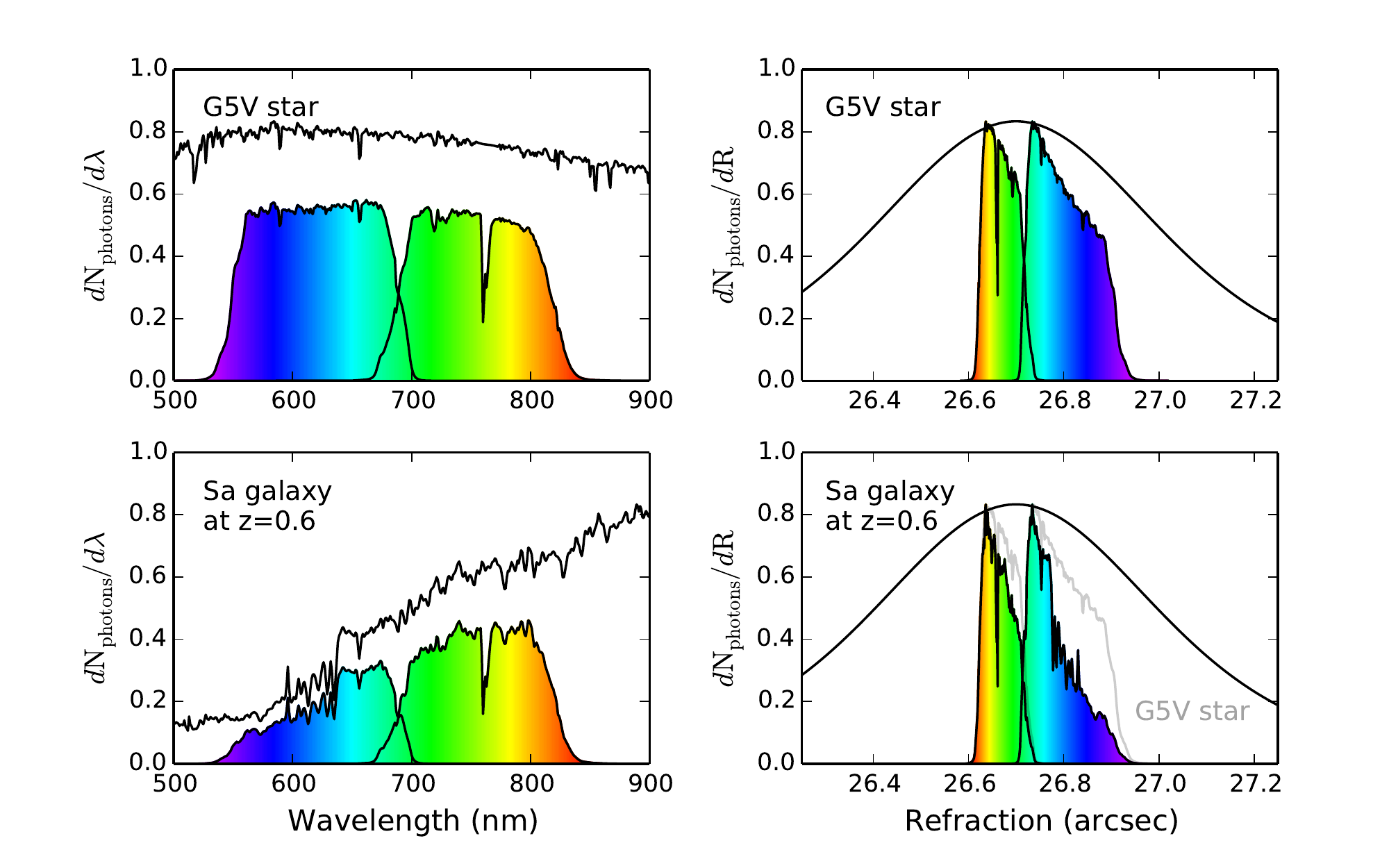}
    \end{center}
    \caption{
      \label{fig:photon_landings} 
      Impact of differential chromatic refraction (DCR) for sample stellar (top) and galactic (bottom) SEDs.
      In the left panels, the black curves show the spectra of a G5V star (top) and an Sa galaxy redshifted to $z=0.6$ (bottom).
      The rainbow-colored areas show the wavelength distribution (with arbitrary normalization) of detected photons from these SEDs after passing through the atmosphere and the LSST filters ($r$- and $i$-band), optics, and sensors.
      In the right panels, we show the amount of refraction for a zenith angle of 35 degrees for the same detected photons, with the same color coding (i.e., colored by wavelength) as the left panels.
      These distributions represent the DCR kernel that is convolved in the zenith direction with the seeing kernel, which is illustrated in the right panels as a $\beta=3.0$ Moffat profile with FWHM of $0.7$ arcseconds (black curve).
      The DCR kernel for the G5V SED is lightly overplotted in the lower right panel to ease comparison.
      The difference between stellar and galactic SEDs therefore leads to different PSFs.
    }
  \end{figure*}

\item %2
  Wavelength dependence of seeing --
  In typical conditions, the dominant component of the PSF for large ground-based telescopes can be traced to changes in refractive index among turbulent cells in the atmosphere. 
The standard theory of atmospheric turbulence predicts that the linear size $\theta$ of the atmospheric convolution kernel (i.e., the seeing) is related to wavelength as $\theta \propto \lambda^{-1/5}$ \citep{Fried66}.  
  For comparison with DCR, Figure~\ref{fig:DCRfilters} also shows a $\lambda^{-2/5}$ relation, which is relevant for comparing PSF second moments and, as we will see in Section \ref{sec:seeing}, is more relevant than the linear seeing for characterizing chromatic shape measurement biases.
  For LSST, the seeing at wavelengths at opposite sides of the $r$-band ($i$-band) differs by about $4.5\%$ ($3.5\%$) (i.e., $9\%$ -- $7\%$ in area) -- independent of the zenith angle\footnote{Note, however, that seeing at a fixed wavelength increases with zenith angle $z_a$ as $\cos^{-3/5}z_a$ due to the increase in airmass.}.
  
\end{enumerate}

These effects can both be described by their impact on a monochromatic reference PSF.
DCR shifts the PSF centroid as a function of wavelength, which manifests as a convolution of the reference PSF by a function that depends on the SED.
Chromatic seeing dilates the PSF as a function of wavelength, which to first order manifests as an SED-dependent dilation of the reference PSF.

Although the effects of DCR could at least partially be mitigated by introducing an atmospheric dispersion corrector (ADC) into the optical path, neither DES nor LSST use such a device.
An ADC would also do nothing to help mitigate chromatic seeing, which we will see can be at least as significant to weak lensing science as DCR.

The potential impact of chromatic effects on the sensitivity and robustness of cosmic shear measurements has been described in five papers.
\citet{Cypriano++10}, \citet{Voigt++12}, and \citet{Semboloni++13} use a range of analytic and computational techniques to study the wavelength dependence of the PSF when it is dominated by the diffraction limit of the telescope, and when the filter bands are several hundred nanometers wide -- i.e., for a space-based survey such as {\it Euclid} \citep{Laureijs++11}.
\citet{Voigt++12} and \citet{Semboloni++13} also study the impact of `color gradients' -- i.e., spatially dependent SEDs in a galaxy.
\citet[hereafter PB12]{Plazas+Bernstein12} study the impact of atmospheric DCR on the first and second zenith-direction moments of images of stars and galaxies with realistic SEDs, using analytic techniques.  
The general conclusions of the above studies are that the biases in shape measurements due to DCR for LSST and due to wavelength dependence of the diffraction limit and galaxy color gradients for {\it Euclid} exceed requirements if not corrected.
The authors explore various methods for correcting the observed biases based on single-color or multi-color photometry and calibration of the bias using multi-colored observations from space.
\citet{Meyers+Burchat14} investigate a power-law dependence of PSF FWHM with wavelength, which can model both a diffraction-limited PSF and chromatic seeing.
To our knowledge, no other published study to date has investigated the impact of chromatic seeing on weak-lensing surveys, though the issue is mentioned briefly in \citet{Weinberg++12}.

In this paper, we estimate the impact of each atmospheric chromatic effect in different filter bands, for ground-based surveys such as the Dark Energy Survey~\citep[DES,][]{DES++07} or LSST, using both analytic and computational techniques.
In Section~\ref{sec:analytics}, we develop the analytic formalism to describe wavelength-dependent PSF effects, and apply this formalism to the specific cases of DCR and chromatic seeing.
In Section~\ref{sec:requirements}, we propagate survey shear measurement requirements into requirements on the parameters that describe chromatic PSF biases; then in Section~\ref{sec:catalog} we show that current and future lensing surveys do not meet these requirements if chromatic effects are left uncorrected.
In Section~\ref{sec:machine_learning}, we show how machine-learning techniques can be applied to the photometry for individual stars and galaxies to infer and account for chromatic biases for each object.
Then, in Section~\ref{sec:model_fitting}, we demonstrate a limitation to the analytically derived chromatic biases similar to the model-fitting or underfitting bias described in the literature, and in 
Section~\ref{sec:psf_corr} we investigate the efficacy of PSF-level corrections.
We identify other potential chromatic effects in Section~\ref{sec:discussion}. 
In Section~\ref{sec:software}, we briefly describe the open-source software tools developed for this work. 
We present our summary and outlook in Section~\ref{sec:conclusions}.

\section{Analytic formalism} %2
\label{sec:analytics}
Weak gravitational lensing is frequently analyzed through its effect on combinations of the second central moments $I_{\mu\nu}$ of a galaxy's surface brightness distribution.
The second central moments of a general surface brightness distribution $I(x,y)$ (applicable to both stars and galaxies, both before and after convolution with the PSF) are given by
\begin{equation}
  \label{eqn:M2}
  I_{\mu \nu} = \frac{1}{f}\int{ I(x,y)(\mu - \bar{\mu})(\nu - \bar{\nu})\dif{x} \dif{y}},
\end{equation}
where $\mu$ and $\nu$ each refer to angular coordinate $x$ or $y$.
The centroids $\bar{\mu}$ and $\bar{\nu}$ and the total flux $f$ of the galaxy are given by
\begin{equation}
  \label{eqn:M1}
  \bar{\mu} = \frac{1}{f}\int{I(x,y)\mu \dif{x} \dif{y}},
\end{equation}
\begin{equation}
  \label{eqn:M0}
  f = \int{I(x,y) \dif{x} \dif{y} }.
\end{equation}
To evaluate the moments on real images, a weight function $w(x, y)$ is required in the integrands of Equations \ref{eqn:M2}-\ref{eqn:M0} to mitigate the effects of noise and the integrals become sums over pixels.
In this paper we primarily focus on galaxies with no color gradients; i.e., we assume that the spatial and wavelength dependence of surface brightness are separable:
\begin{equation}
\label{eqn:no_color_gradients}
  I(x,y,\lambda) = I(x,y) S(\lambda),
\end{equation}
where $I(x,y,\lambda)$ is the wavelength-dependent surface brightness distribution and $S(\lambda)$ is the SED of the galaxy.
This assumption allows us to absorb the wavelength dependence of the PSF into an effective PSF for each potential SED:
\begin{equation}
\label{eqn:PSFeff}
  \mathrm{PSF}_\mathrm{eff}(x,y) \propto \int{\mathrm{PSF}(x,y,\lambda) S(\lambda) F(\lambda) \dif{\lambda}},
\end{equation}
where $F(\lambda)$ is the total response function\footnote{Note that we use the total response function at an airmass of 1.2 throughout this paper, independent of zenith angle.} and the normalization of the effective PSF is such that its integral over all space is 1.
Hereafter, we will omit the word `effective' when the context makes clear the distinction between the full wavelength-dependent PSF ($\mathrm{PSF}(x,y,\lambda)$, which is independent of any SED or throughput function) and a wavelength-independent effective PSF ($\mathrm{PSF}_\mathrm{eff}(x,y)$, which requires that an SED and throughput function be specified).

In general, four specific cases of second moments are relevant for estimation of a pre-PSF-convolution galaxy shape using a stellar-estimated PSF:
\begin{enumerate}
  \item %1
  The moments of the lensed galaxy before convolution with the PSF: $I_{\mu \nu}^\mathrm{gal}$.
  \item %2
  The moments of the galactic PSF: $I_{\mu \nu}^\mathrm{PSF,g}$.
  \item %3
  The moments of the observed galaxy image\footnote{This relation holds exactly for unweighted second moments (i.e., when $w(x, y) = 1$).  However, this assumption breaks down when using more general weight functions or for model-fitting shape measurement approaches, in which the model being fit acts like a weight function.  See Sec.~\ref{sec:model_fitting}.}:
  \begin{equation}
    \label{eqn:second_moments_add}
    I_{\mu \nu}^\mathrm{obs} = I_{\mu \nu}^\mathrm{gal} + I_{\mu \nu}^\mathrm{PSF,g}.
  \end{equation}
  \item %4
  The moments of the stellar PSF: $I_{\mu \nu}^\mathrm{PSF,*}$.
\end{enumerate}
Note that we use the superscript $\mathrm{g}$ rather than $\mathrm{gal}$ in $I_{\mu\nu}^\mathrm{PSF,g}$ to emphasize that this symbol refers to the second moment of the (galactic) PSF, not the second moment of the galaxy ($I_{\mu \nu}^\mathrm{gal}$).
The quantity of interest for weak lensing is $I_{\mu \nu}^\mathrm{gal}$, but only $I_{\mu \nu}^\mathrm{obs}$ and $I_{\mu \nu}^\mathrm{PSF,*}$ are directly observable.

Important combinations of second central moments are the second-moment squared radius $r^2$
(e.g., $r_{\rm PSF}^2$ for the PSF and $r_{\rm gal}^2$ for the galaxy before convolution with the PSF)
and two complex ellipticities $\boldsymbol{\chi} = \chi_1 + \mathrm{i}\chi_2$ and $\boldsymbol{\epsilon} = \epsilon_1 + \mathrm{i}\epsilon_2$:
\begin{equation}
  \label{eqn:r2}
  r^2 = I_{xx} + I_{yy},
\end{equation}
\begin{equation}
  \label{eqn:chi1}
  \chi_1 = \frac{I_{xx} - I_{yy}}{I_{xx} + I_{yy}},
\end{equation}
\begin{equation}
  \label{eqn:chi2}
  \chi_2 = \frac{2 I_{xy}}{I_{xx} + I_{yy}},
\end{equation}
\begin{equation}
  \label{eqn:eps1}
  \epsilon_1 = \frac{I_{xx} - I_{yy}}{I_{xx} + I_{yy} + 2\sqrt{I_{xx}I_{yy}-I_{xy}^2}},
\end{equation}
\begin{equation}
  \label{eqn:eps2}
  \epsilon_2 = \frac{2 I_{xy}}{I_{xx} + I_{yy} + 2\sqrt{I_{xx}I_{yy}-I_{xy}^2}}.
\end{equation}
Which definition of ellipticity is more convenient depends on the context; we will use both in this paper.
An object with perfectly elliptical isophotes and ratio $q$ of minor to major axes ($0 \le q \le 1$) has $\chi$- and $\epsilon$-ellipticity magnitudes equal to
\begin{equation}
  \label{eqn:q_to_chi}
  |\boldsymbol{\chi}| = \frac{1-q^2}{1+q^2},
\end{equation}
\begin{equation}
  \label{eqn:q_to_eps}
  |\boldsymbol{\epsilon}| = \frac{1-q}{1+q}.
\end{equation}
A galaxy's apparent (lensed) ellipticity $\boldsymbol{\chi^{(a)}}$ or $\boldsymbol{\epsilon^{(a)}}$ is related to its intrinsic (unlensed) ellipticity $\boldsymbol{\chi^{(i)}}$ or $\boldsymbol{\epsilon^{(i)}}$ in the presence of gravitational lensing shear $\boldsymbol{\gamma} = \gamma_1 + \mathrm{i} \gamma_2$ and convergence $\kappa$ via
\begin{equation}
  \label{eqn:applyshear_chi}
  \boldsymbol{\chi^{(a)}} = \frac{\boldsymbol{\chi^{(i)}} + 2 \boldsymbol{g} + \boldsymbol{g}^2 {\boldsymbol{\overline\chi^{(i)}}}}{1+|\boldsymbol{g}|^2+2 \Re(\boldsymbol{g}{\boldsymbol{\overline\chi^{(i)}}})},
\end{equation}
\begin{equation}
  \label{eqn:applyshear_eps}
  \boldsymbol{\epsilon^{(a)}} = \frac{\boldsymbol{\epsilon^{(i)}} + \boldsymbol{g}}{1+\boldsymbol{\overline{g}}\boldsymbol{\epsilon^{(i)}}},
\end{equation}
where $\boldsymbol{g} = \boldsymbol{\gamma}/(1-\kappa)$ is the reduced shear and the overbar indicates complex conjugation \citep{Schneider+Seitz95, Seitz+Schneider97}.
Under the assumption that intrinsic galaxy ellipticities are isotropically distributed, the mean of the sheared ellipticities is related to the reduced shear by
\begin{equation}
  \left\langle \boldsymbol{\chi^{(a)}} \right\rangle \approx 2 \boldsymbol{g},
\end{equation}
\begin{equation}
  \left\langle \boldsymbol{\epsilon^{(a)}} \right\rangle = \boldsymbol{g},
\end{equation}
where the expectation value is exact for $\epsilon$-ellipticity but ignores a $\sim$ 10\% correction that depends on the distribution of intrinsic ellipticities for $\chi$-ellipticity.

To characterize chromatic PSF effects, it is convenient to define the difference between second central moments of galactic and stellar PSFs:
\begin{equation}
  \label{eqn:deltaI}
  \Delta I_{\mu \nu}^\mathrm{PSF} \equiv I_{\mu \nu}^\mathrm{PSF,g} - I_{\mu \nu}^\mathrm{PSF,*}.
\end{equation}
Substituting the observable stellar PSF for the unobservable galactic PSF shifts the inferred pre-seeing galactic moments:
$I_{\mu \nu}^\mathrm{gal} \rightarrow I_{\mu \nu}^\mathrm{gal}+\Delta I_{\mu \nu}^\mathrm{PSF}$.
This change propagates into $\chi_1$ and $\chi_2$ (the propagation into $\epsilon_1$ and $\epsilon_2$ is also possible, but less convenient) as
\begin{equation}
  \label{eqn:chi1prop}
  \begin{split}
    \chi_1& \rightarrow \frac{(I_{xx}^\mathrm{gal}+\Delta I_{xx}^\mathrm{PSF})-(I_{yy}^\mathrm{gal}+\Delta I_{yy}^\mathrm{PSF})}{(I_{xx}^\mathrm{gal}+\Delta I_{xx}^\mathrm{PSF})+(I_{yy}^\mathrm{gal}+\Delta I_{yy}^\mathrm{PSF})}\\
    & \simeq \chi_1\left(1-\frac{\Delta I_{xx}^\mathrm{PSF}+\Delta I_{yy}^\mathrm{PSF}}{r^2_\mathrm{gal}}\right) + \frac{\Delta I_{xx}^\mathrm{PSF}-\Delta I_{yy}^\mathrm{PSF}}{r^2_\mathrm{gal}} + \mathcal{O}\left(\Delta I\right)^2,\\
  \end{split}
\end{equation}
\begin{equation}
  \label{eqn:chi2prop}
  \begin{split}
    \chi_2& \rightarrow \frac{2(I_{xy}^\mathrm{gal}+\Delta I_{xy}^\mathrm{PSF})}{(I_{xx}^\mathrm{gal}+\Delta I_{xx}^\mathrm{PSF})+(I_{yy}^\mathrm{gal}+\Delta I_{yy}^\mathrm{PSF})}\\
    & \simeq \chi_2\left(1-\frac{\Delta I_{xx}^\mathrm{PSF}+\Delta I_{yy}^\mathrm{PSF}}{r^2_\mathrm{gal}}\right) + \frac{2\Delta I_{xy}^\mathrm{PSF}}{r^2_\mathrm{gal}} + \mathcal{O}\left(\Delta I\right)^2.\\
  \end{split}
\end{equation}
These formulae are essentially the same as Equation 13 of \citet{Paulin-Henriksson++08}, but framed in terms of errors in the second central moments of the PSF instead of errors in the PSF size and ellipticity.

We follow the literature~\citep{Heymans++05} and parameterize the bias in the reduced shear in terms of multiplicative and additive terms,
\begin{equation}
  \label{eqn:shearbias}
\hat{g}_i=g_i (1+m_i)+c_i,\quad i=1,2,
\end{equation}
where $\hat{\boldsymbol{g}}$ is the estimator for the true reduced shear $\boldsymbol{g}$ and we have assumed that $\hat{g}_1$ ($\hat{g}_2$) is independent of $g_2$ ($g_1$).
The shear calibration parameters are then given by
\begin{equation}
  \label{eqn:m1m2}
  m_1=m_2=\frac{-\left(\Delta I_{xx}^\mathrm{PSF}+\Delta I_{yy}^\mathrm{PSF}\right)}{r^2_\mathrm{gal}},
\end{equation}

\begin{equation}
  \label{eqn:c1}
  c_1=\frac{\Delta I_{xx}^\mathrm{PSF}-\Delta I_{yy}^\mathrm{PSF}}{2 r^2_\mathrm{gal}},
\end{equation}

\begin{equation}
  \label{eqn:c2}
  c_2=\frac{\Delta I_{xy}^\mathrm{PSF}}{r^2_\mathrm{gal}}.
\end{equation}

\subsection{Differential Chromatic Refraction}
\label{sec:dcr}
The refraction angle (i.e., the change in the observed zenith angle due to refraction) $R(\lambda; z_a)$ can be expressed as the product of a factor that depends on the wavelength $\lambda$ of the photon and a factor that depends on the zenith angle $z_a$:
\begin{equation}
  \label{eqn:Rlamza}
  R(\lambda; z_a) =  h(\lambda) \tan z_a,
\end{equation}
where $h(\lambda)$ depends implicitly on the air pressure, temperature, and partial pressure of water vapor in the telescope dome, and can be obtained from formulae given by \citet{Edlen53} and \citet{Coleman++60}.
For monochromatic sources, the dominant effect is to move the apparent position of the object.
For sources that are not monochromatic (i.e., all real sources), with a wavelength distribution of detected photons (i.e., the product of the source photon distribution and the total system throughput) given by $p_{\lambda}(\lambda)$, the variation in displacement of photons with different wavelengths introduces a convolution kernel $k(R)$ in the zenith direction that can be written in terms of the inverse $\lambda(R; z_a)$ of Equation \ref{eqn:Rlamza}:
\begin{equation}
  \label{eqn:convker}
  k(R) = \frac{p_\lambda(\lambda(R; z_a)) \left|\od{\lambda}{R}\right|}{\int{p_\lambda(\lambda) \dif{\lambda}}}.
\end{equation}
This kernel can largely be characterized in terms of its first moment $\bar{R}$ and its second central moment, or variance, $V$:
\begin{equation}
  \label{eqn:Rbar}
  \begin{split}
    \bar{R} & = \int{k(R) R \dif{R} } \\
            & = \frac{\int{p_\lambda(\lambda)R(\lambda; z_a)\dif{\lambda} }}{\int{p_\lambda(\lambda) \dif{\lambda}}},
  \end{split}
\end{equation}
\begin{equation}
  \label{eqn:V}
  \begin{split}
    V & = \int{k(R) (R - \bar{R})^2 \dif{R} } \\
      & = \frac{\int{p_\lambda(\lambda)(R(\lambda; z_a) - \bar{R})^2\dif{\lambda} }}{\int{p_\lambda(\lambda) \dif{\lambda}}}.
  \end{split}
\end{equation}

\subsubsection{DCR First-Moment Shifts}
\label{sec:DCRfirstmoment}

In a single epoch, the shifts in star and galaxy PSF centroids due to DCR do not introduce galaxy shape measurement bias.
However, if measurements on a given patch of sky are made from a stack of observations that span a range in zenith and parallactic angles, then the resulting ensemble of relative shifts $\Delta \bar{R} \equiv \bar{R}^\mathrm{gal} - \bar{R}^\mathrm{*}$, each of which may point in a different direction in celestial coordinates, will lead to misregistration of the galaxy center among the different epochs, which can lead to shear biases if not taken into account.
For an object whose celestial coordinates are given by declination $\delta$ and right ascension $\alpha$, the relative centroid shifts in a small patch of sky (so that the coordinates can be approximated as rectilinear) are
\begin{equation}
  \label{eqn:RA}
  (\cos \delta) \Delta \alpha = \Delta \bar{R}_{45} \tan z_a \sin q
\end{equation}
and
\begin{equation}
  \label{eqn:dec}
\Delta \delta = \Delta \bar{R}_{45} \tan z_a \cos q,
\end{equation}
where $\Delta \bar{R}_{45}$ is the difference in the first moments of the refraction of a star and galaxy at a zenith angle $z_a$ of 45 degrees and $q$ is the parallactic angle of the galaxy (i.e., the position angle of the zenith measured from North going East).
As $z_a$ and $q$ vary with epoch over the course of the survey, so will these centroid shifts.

Assuming for the moment that all stars have the same SED, the means over epochs of the shifts, $\langle(\cos \delta)\Delta \alpha\rangle_\mathrm{epochs}$ and $\langle \Delta \delta\rangle_\mathrm{epochs}$, indicate the 2D centroid shift, relative to the stars, for the stacked galaxy image when the individual images are registered from the positions of the stars.
This is very nearly still the case if the stars are given a realistic distribution of SEDs because the number of star positions being averaged in the registration process is large.
For each galaxy, the second central moment $I_{\mu \nu}^\mathrm{stack}$ of the stacked surface brightness distribution is equal to the sum of the second central moment of the single-epoch surface brightness distribution and the second central moment of the 2D distribution of centroid shifts:
\begin{equation}
  I_{\mu \nu}^\mathrm{stack} = I_{\mu \nu}^\mathrm{single\,epoch} + \langle(\mu - \bar{\mu})( \nu - \bar{\nu})\rangle_\mathrm{epochs},
  \label{eqn:misregistration_moments}
\end{equation}
where $\mu$ and $\nu$ are either $(\cos \delta) \Delta \alpha$ or $\Delta \delta$, and $\bar{\mu}$ and $\bar{\nu}$ are the means of these quantities over epochs.
Since the last term in Equation \ref{eqn:misregistration_moments} enters into the observed galaxy second moment in exactly the same way as the PSF, we can interpret it as an error in the PSF second moments:
\begin{equation}
  \Delta I_{\mu \nu}^\mathrm{PSF} = \langle(\mu-\bar{\mu})(\nu-\bar{\nu})\rangle_\mathrm{epochs}.
\end{equation}
This second-moment error can then be propagated into a shear measurement bias through Equations~\ref{eqn:m1m2}-\ref{eqn:c2}.

\begin{figure}
  \begin{center}
    \epsscale{1.1}
    \plotone{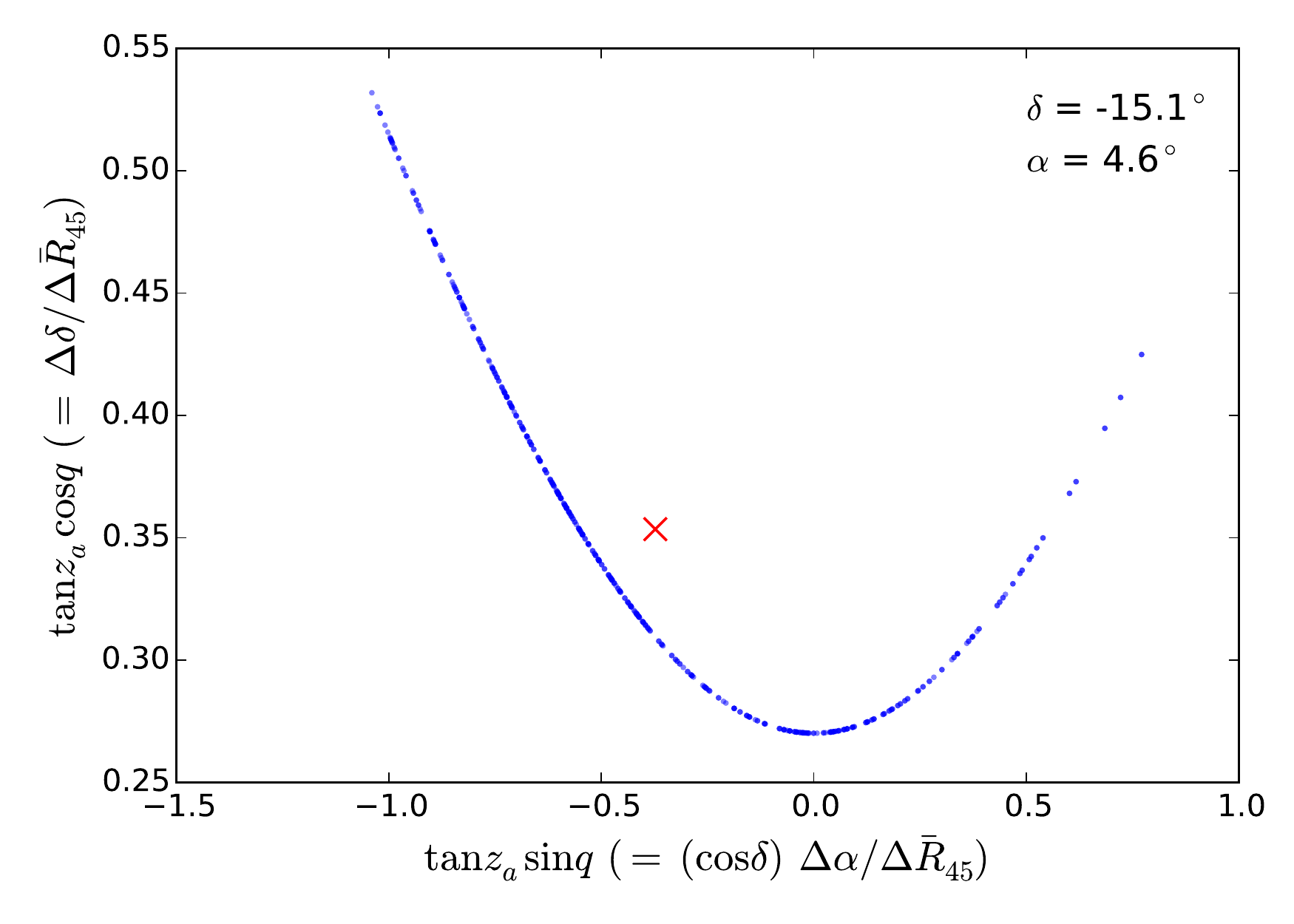}
  \end{center}
  \caption{The 2D distribution of $\tan z_a \sin q=(\cos \delta) \Delta \alpha / \Delta \bar{R}_{45}$ and $\tan z_a \cos q=\Delta \delta / \Delta \bar{R}_{45}$ for one particular field in the LSST Operations Simulation Run 3.61~\citep{Delgado++06}.
  The distribution follows a curve due to geometric constraints.
  The centroid of the distribution is indicated by the red `\red{$\times$}'.}
  \label{fig:epoch_variance}
\end{figure}

\begin{figure}
  \begin{center}
    \epsscale{1.1}
    \plotone{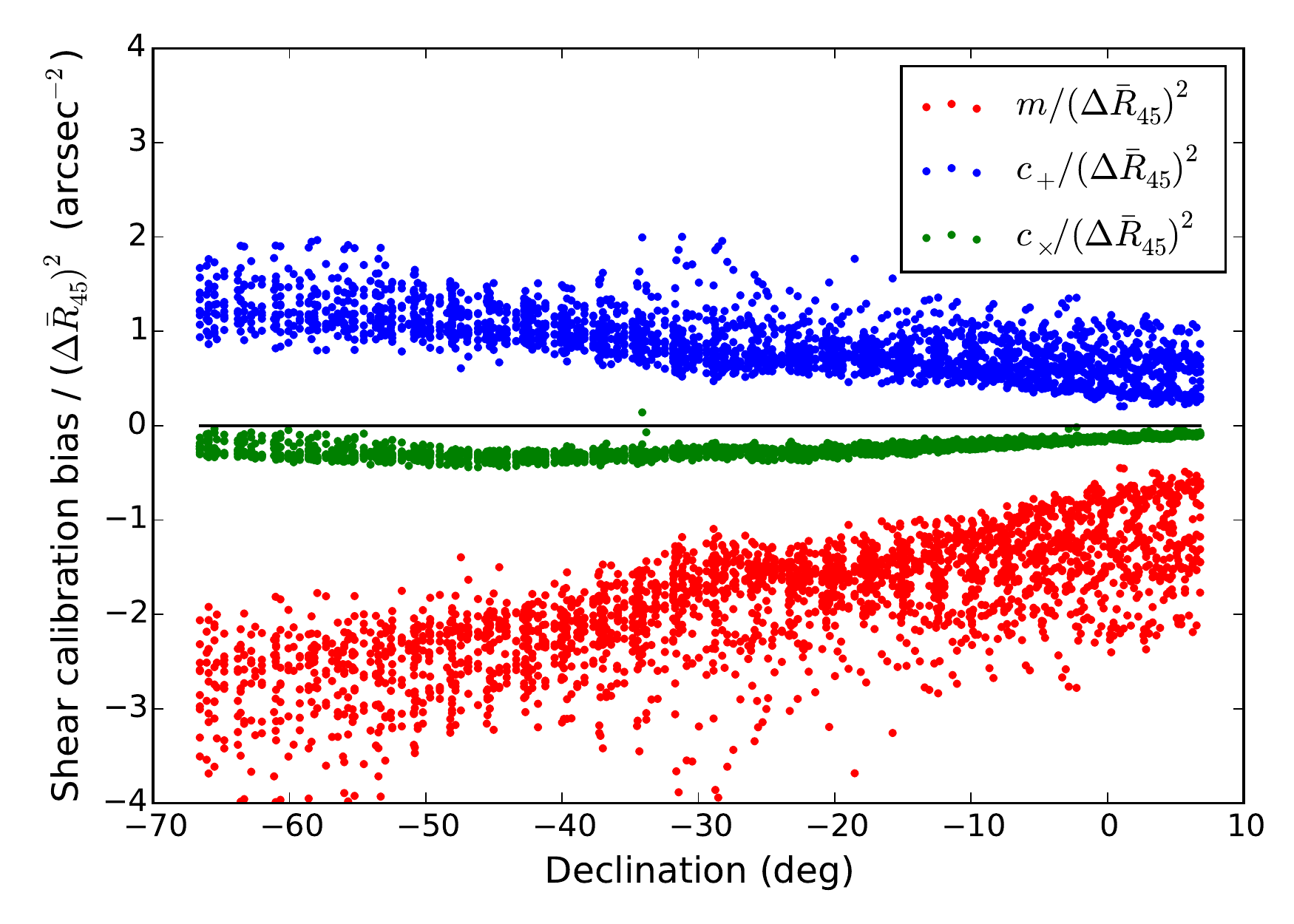}
  \end{center}
  \caption{SED-independent estimates of chromatic biases due to DCR-induced misregistration for each wide-fast-deep LSST field in the Operations Simulation Run 3.61.
  Each plotting symbol in this figure corresponds to a single LSST field.
  To obtain $m$ or $c_{+,\times}$ for a particular galaxy SED and redshift multiply the value on the vertical axis by the value of $(\Delta \bar{R})^2$ from the top panel of Figure~\ref{fig:bias_panel}.
The results here correspond to a possible set of observations for a ten-year LSST survey.}
  \label{fig:misregistration_bias}
\end{figure}

To estimate the impact of misregistration bias we use a simulation that predicts the distribution of zenith and parallactic angles at which each patch of sky is observed for a possible LSST ten-year survey\footnote{Run 3.61 of the LSST Operations Simulator \citep{Delgado++06}.}.
From the zenith and parallactic angles in this simulation, we use Equations \ref{eqn:RA} and \ref{eqn:dec} to investigate the distributions of
$ \tan z_a \sin q = (\cos \delta) \Delta \alpha / \Delta \bar{R}_{45}$ and
$\tan z_a \cos q = \Delta \delta / \Delta \bar{R}_{45}$ for each LSST field. (Figure \ref{fig:epoch_variance} shows an example for one such field.)
By dividing by $\Delta \bar{R}_{45}$, we factor out the bias dependence on the SEDs of the stars and galaxies.
We then compute $\langle(\mu - \bar{\mu})(\nu-\bar{\nu})\rangle_\mathrm{epochs} / (\Delta \bar{R}_{45})^2 = \Delta I_{\mu \nu}^\mathrm{PSF} / (\Delta \bar{R}_{45})^2$ and finally the scaled shear bias parameters $m / (\Delta \bar{R}_{45})^2$ and $c / (\Delta \bar{R}_{45})^2$.
For the shear bias parameter estimates, we assume a typical galaxy second-moment squared radius of $(0.3\arcsec)^2$, which we will motivate further in Section \ref{sec:requirements}.
Figure \ref{fig:misregistration_bias} shows the resulting shear biases plotted as a function of field declination.
The multiplicative bias is negative, implying that misregistration tends to enlarge the stacked image along both axes to make the galaxy look rounder.
The additive bias in the `$+$' direction -- i.e., along the $\alpha$ and $\delta$ axes -- is positive, which is consistent with the misregistration enlarging the stacked galaxy image primarily along the RA axis as shown for example in Figure \ref{fig:epoch_variance}.
The additive bias in the `$\times$' direction is nearly zero, which indicates that the distribution of hour angles for observing a target at a given declination is nearly symmetric about 0h.
While these results are specifically estimated for LSST, the impact of misregistration bias for DES is likely similar as DES is limited to using larger galaxies, but typically observes at larger zenith angles.

\subsubsection{DCR Second-Moment Shifts}

In contrast to DCR first-moment shifts, differences in star and galaxy PSF second moments induce shear biases even in individual epochs.
If the $y$ direction is defined to be toward the zenith\footnote{Note that this differs from the common convention of having $+y$ point North.}, the effect of DCR is to take $I_{yy}^\mathrm{PSF} \rightarrow I_{yy}^\mathrm{PSF} + V$, but leave $I_{xx}^\mathrm{PSF}$ and $I_{xy}^\mathrm{PSF}$ unchanged.
The difference between stellar and galactic SEDs leads to $\Delta I^\mathrm{PSF,DCR}_{yy} = \Delta V \equiv V^\mathrm{gal} - V^\mathrm{*} \ll r^2_\mathrm{PSF}$ and $\Delta I^\mathrm{PSF,DCR}_{xy}=\Delta I^\mathrm{PSF,DCR}_{xx}=0$.
Inserting $\Delta I_{\mu \nu}^\mathrm{PSF,DCR}$ into Equations \ref{eqn:m1m2}-\ref{eqn:c2} reveals the multiplicative and additive shear calibration biases due to DCR-induced second-moment shifts:
\begin{equation}
  \label{eqn:m1m2_DCR}
  m_1^\mathrm{DCR} = m_2^\mathrm{DCR} = -\frac{\Delta V}{r^2_\mathrm{gal}},
\end{equation}
\begin{equation}
  \label{eqn:c1_DCR}
  c_1^\mathrm{DCR} = -\frac{\Delta V}{2 r^2_\mathrm{gal}},
\end{equation}
\begin{equation}
  \label{eqn:c2_DCR}
  c_2^\mathrm{DCR} = 0.
\end{equation}

\subsection{Wavelength Dependence of Seeing}
\label{sec:seeing}
Kolmogorov turbulence in the atmosphere leads to a PSF linear size $\theta$ that scales like $\theta \propto \lambda^{-1/5}$, introducing a second atmospheric wavelength dependence to the PSF.
Assuming that this wavelength dependence corresponds to linear isotropic dilation or contraction\footnote{This is the case for Kolmogorov turbulence in the long-exposure limit.} of a fiducial atmospheric PSF (say, the monochromatic PSF at wavelength $\lambda_0$), then the second central moments at a given wavelength can be written
\begin{equation}
  \label{eqn:seeing_M2_lambda}
  I^{\mathrm{PSF},\lambda}_{\mu\nu} = I^{\mathrm{PSF},\lambda_0}_{\mu\nu} \left(\frac{\lambda}{\lambda_0}\right)^{-2/5}.
\end{equation}
The SED-weighted second moments are hence
\begin{equation}
  \label{eqn:seeing_M2_SED}
  I^\mathrm{PSF}_{\mu\nu} = I^{\mathrm{PSF},\lambda_0}_{\mu\nu} \frac{\int{p_\lambda(\lambda)\left(\lambda/\lambda_0\right)^{-2/5}\dif{\lambda}}}{\int{p_\lambda(\lambda)\dif{\lambda}}};
\end{equation}
i.e., for a given reference wavelength and associated monochromatic reference PSF, the second central moments of the effective PSF generated from any particular SED are related to the second central moments of the reference PSF by a fixed multiplier.
Since the second-moment squared radius is a linear combination of second central moments, we can use it as the scale factor relating stellar and galactic PSFs:
\begin{equation}
  I^\mathrm{PSF,g}_{\mu\nu} = I^\mathrm{PSF,*}_{\mu\nu} \frac{r^2_\mathrm{PSF,g}}{r^2_\mathrm{PSF,*}},
\end{equation}
where the two second-moment squared radii implement a simple rescaling of the second-moment matrix.
Some algebra then reveals that the differences in PSF second moments between stars and galaxies are
\begin{equation}
  \label{eqn:delta_M2_seeing}
  \Delta I^\mathrm{PSF, seeing}_{\mu\nu} = I^\mathrm{PSF,*}_{\mu\nu} \frac{\Delta r^2_\mathrm{PSF}}{r^2_\mathrm{PSF,*}},
\end{equation}    
where
\begin{equation}
  \label{eqn:delta_r2}
  \Delta r^2_\mathrm{PSF} = r^2_\mathrm{PSF,g} - r^2_\mathrm{PSF,*}.
\end{equation}

Inserting Equation \ref{eqn:delta_M2_seeing} into Equations \ref{eqn:m1m2}-\ref{eqn:c2} gives the multiplicative and additive shear calibration biases due to chromatic seeing:
\begin{equation}
  \label{eqn:m1m2_seeing}
  m^\mathrm{seeing}_1 = m^\mathrm{seeing}_2 = -\frac{r^2_\mathrm{PSF}}{r^2_\mathrm{gal}} \frac{\Delta r^2_\mathrm{PSF}}{r^2_\mathrm{PSF}},
\end{equation}
\begin{equation}
  \label{eqn:c1_seeing}
  c^\mathrm{seeing}_1 = \frac{I_{xx}^\mathrm{PSF,*} - I_{yy}^\mathrm{PSF,*}}{2r^2_\mathrm{gal}}\frac{\Delta r^2_\mathrm{PSF}}{r^2_\mathrm{PSF}} = \frac{\chi_1^\mathrm{PSF}}{2} \frac{r^2_\mathrm{PSF}}{r^2_\mathrm{gal}} \frac{\Delta r^2_\mathrm{PSF}}{r^2_\mathrm{PSF}},
\end{equation}
\begin{equation}
  \label{eqn:c2_seeing}
  c^\mathrm{seeing}_2 = \frac{I_{xy}^\mathrm{PSF,*}}{r^2_\mathrm{gal}}\frac{\Delta r^2_\mathrm{PSF}}{r^2_\mathrm{PSF}} = \frac{\chi_2^\mathrm{PSF}}{2} \frac{r^2_\mathrm{PSF}}{r^2_\mathrm{gal}} \frac{\Delta r^2_\mathrm{PSF}}{r^2_\mathrm{PSF}}.
\end{equation}

Note that we have explicitly chosen not to cancel the $r^2_\mathrm{PSF}$ factors that appear in both the numerator and denominator of these expressions.
This is because the ratio $\Delta r^2_\mathrm{PSF}/r^2_\mathrm{PSF}$ -- i.e., the fractional difference in PSF sizes -- is convenient to preserve, as it does not depend on the current atmospheric conditions.
This is in contrast to the absolute PSF size difference $\Delta r^2_\mathrm{PSF}$, which scales with the square of the seeing.

\subsection{DCR and Chromatic Seeing Together}
\label{sec:both}
Of course, a real atmospheric PSF contains chromatic effects from both DCR and wavelength-dependent seeing.
We treat both these effects as wavelength-dependent perturbations to a fiducial PSF -- for example, the monochromatic PSF at some reference wavelength.
The two perturbations -- a wavelength-dependent shift in the case of DCR and a wavelength-dependent dilation (or contraction) with respect to the centroid in the case of chromatic seeing -- do not commute.
Therefore, it is important to identify the order in which the effects apply.
The correct order is a dilation followed by a shift, since, in the opposite order, the dilation about the unperturbed PSF centroid would exaggerate the overall shift.

\section{Requirements on Chromatic Bias Parameters} %3
\label{sec:requirements}
The tolerance of a given survey to non-zero shear bias depends on its statistical power, which in turn is determined largely by the survey sky area $A$, effective number density of galaxies, $n_\mathrm{eff}$, and median redshift $z_m$ \citep{Amara+Refregier08}.
In Table~\ref{table:surveys}, we list these survey characteristics for DES and LSST.

\begin{deluxetable}{lccccc}[!]
  \tablecaption{\label{table:surveys} Survey characteristics and shear-bias requirements.}
  \tablehead{
    \colhead{Survey} &
    \colhead{Area} &
    \colhead{$n_\mathrm{eff}$} &
    \colhead{$z_m$} &
    \colhead{$|\langle m\rangle|_\mathrm{max}$} &
    \colhead{$\mathrm{Var}(c)_\mathrm{max}$}
    }
  \startdata
    DES    & 5000\tablenotemark{a}  & 12\tablenotemark{a} & 0.68\tablenotemark{a} & 0.008 & $6.0 \times 10^{-7}$ \\
    LSST   & 18000\tablenotemark{b} & 30\tablenotemark{c} & 0.82\tablenotemark{c} & 0.003 & $1.8 \times 10^{-7}$
  \enddata
  \tablenotetext{a}{\citet{DES++07}}
  \tablenotetext{b}{\citet{LSSTSR}}
  \tablenotetext{c}{\citet{Chang++13}}
  \tablecomments{
    Specifications for the DES and LSST surveys and requirements on shear bias parameters.
    The survey area is given in square degrees.
    The parameter $n_\mathrm{eff}$ is the effective number of galaxies per square arcminute, and $z_m$ is the median redshift of these galaxies.
    The multiplicative shear calibration parameter $m$ is defined in Eq.~\ref{eqn:shearbias}.
    The `requirements' $|\langle m\rangle|_\mathrm{max}$ correspond to the values at which the {\it combined} uncertainties from {\it all} multiplicative systematic effects will equal the expected statistical uncertainty of each weak-lensing survey.
    The `requirements' $\mathrm{Var}(c)_\mathrm{max}$ indicate the values at which the {\it combined} uncertainties from {\it all} additive systematic effects could (but will not necessarily) equal the expected statistical uncertainty of each weak-lensing survey.
  }
\end{deluxetable}

The choice of `requirements' on the shear bias parameters $m$ and $c$ is somewhat arbitrary.
Certainly the contribution of any one systematic effect to the uncertainty on cosmological parameters extracted from the survey should be less than the expected statistical uncertainty -- but how much less, given that a yet-undetermined number of systematic effects may contribute?
Rather than choosing an arbitrary fraction of the statistical uncertainty, we will express requirements as the full equivalent statistical uncertainty.
Therefore, we must strive to keep individual systematic effects, such as those due to wavelength-dependent PSFs, well below these values.
We stress that it is the \emph{prior uncertainty} on the shear bias parameters -- not the actual value of the biases -- that leads to systematic uncertainties, since if the biases are known they can be removed from the analysis.

\begin{deluxetable}{lllll}[!]
  \tablecaption{\label{table:sizes} PSF and galaxy sizes.}
  \tablehead{
    \colhead{Survey} &
    \colhead{$\mathrm{FWHM}_\mathrm{PSF}$} &
    \colhead{$r^2_\mathrm{PSF}$\tablenotemark{a}} &
    \colhead{$i$-lim} &
    \colhead{$r^2_\mathrm{gal}$\tablenotemark{b}}
  }
  \startdata
    DES    & $0.8\arcsec$\tablenotemark{d} & $(0.8\arcsec)^2$ & 24\tablenotemark{c}   & $(0.4\arcsec)^2$ \\
    LSST   & $0.7\arcsec$\tablenotemark{e} & $(0.7\arcsec)^2$ & 25.3\tablenotemark{d} & $(0.3\arcsec)^2$
  \enddata
  \tablenotetext{a}{Computed from $\mathrm{FWHM}_\mathrm{PSF}$ assuming a Moffat profile with parameter $\beta=3.0$.}
  \tablenotetext{b}{Mode of the $\sqrt{r^2_\mathrm{gal}}$ distribution in \textsc{CatSim} for galaxies with the specified magnitude limit near the peak of the source-galaxy redshift distribution.}
  \tablenotetext{c}{\citet{DES++07}}
  \tablenotetext{d}{\citet{LSSTSB}}
  \tablecomments{
    Typical expected PSF and galaxy sizes and limiting magnitudes ($i$-lim) for galaxies in weak-lensing analyses in the DES and LSST surveys.
    For the PSF, we quote the full width at half maximum ($\mathrm{FWHM}_\mathrm{PSF}$) and second-moment squared radius ($r^2_\mathrm{PSF}$).
    For the galaxies, we quote the second-moment squared radius ($r^2_\mathrm{gal}$).
  }
\end{deluxetable}

Following PB12, we list requirements on multiplicative bias based on \citet{Huterer++06}, who compute the fractional increase in the uncertainty on the (constant) dark energy equation of state parameter $w$ as a function of the prior uncertainty in $m$ for both DES and LSST.
As just described, we list in Table~\ref{table:surveys} the requirement on our knowledge of $m$ such that the uncertainty in $w$ is degraded to no more than $\sqrt{2}$ times its purely statistical uncertainty, which results in $|\langle m \rangle|_\mathrm{max} = 0.003$ $(0.008)$ for LSST (DES).

In contrast, \citet{Amara+Refregier08} give a fitting formula in $A$, $n_\mathrm{eff}$, and $z_m$ for the requirement on $m$ such that the uncertainty in $w$ is degraded by only $\sim15\%.$
As a result, the DES and LSST requirements on $m$ inferred from the \citet{Amara+Refregier08} formula are about a factor of three smaller than the values presented in Table~\ref{table:surveys}.
Under different assumptions for how well systematic biases need to be controlled, \citet{Massey++13} and \citet{Cropper++13} conclude that for the {\it Euclid} survey (which will have statistical power similar to LSST) $m$ should be kept below about $0.002$, which is consistent at the $\sim 50\%$ level with the requirement we list in Table~\ref{table:surveys} for LSST.

In contrast to multiplicative biases, errors on cosmological parameters due to additive shear calibration biases are more challenging to estimate, primarily because additive biases influence shear two-point functions not only through their mean and variance but also through their angular covariance.
In fact, the contribution to the additive bias that does not depend on angle -- i.e., the mean additive bias -- affects angular power spectra only at $\ell=0$ and hence does not affect cosmological parameter constraints derived from angular power spectra.
Additive bias due to chromatic effects could be anisotropic -- for example, due to the difference in clustering among red and blue galaxies.
In Section~\ref{sec:additive_biases}, we investigate the degree to which residual additive biases after correction are correlated.

However, even without knowledge of their full correlation function, it is possible to constrain the effects of additive shear biases given only their variance.
In Appendix~\ref{sec:addreq}, we show that an additive bias variance less than $1.8 \times 10^{-7}$ ($6.0 \times 10^{-7}$) is sufficient though not necessary to keep systematic uncertainties on $w$ from exceeding the statistical uncertainties for LSST (DES).
To simplify the following discussion we will refer to these numbers, listed in Table~\ref{table:surveys}, as requirements on the variance of $c$.

\begin{deluxetable*}{lcccccc}
  \tablecaption{\label{table:chromatic_requirements} Requirements on chromatic bias parameters. }
  \tablehead{
    \colhead{Survey} &
    \colhead{$\langle(\Delta \bar{R}_{45})^2\rangle_\mathrm{SEDs}$} &
    \colhead{$\mathrm{Var}((\Delta \bar{R}_{45})^2)_\mathrm{SEDs}$} &
    \colhead{$|\langle\Delta V\rangle_\mathrm{SEDs}|$} &
    \colhead{$\mathrm{Var}(\Delta V)_\mathrm{SEDs}$} &
    \colhead{$|\langle\Delta r^2_\mathrm{PSF}/r^2_\mathrm{PSF}\rangle_\mathrm{SEDs}|$} &
    \colhead{$\mathrm{Var}(\Delta r^2_\mathrm{PSF}/r^2_\mathrm{PSF})_\mathrm{SEDs}$}
  }
  \startdata
    DES  & $4.0 \times 10^{-3}\,\mathrm{arcsec}^2$ & $6.0 \times 10^{-7}\,\mathrm{arcsec}^4$ & $1.3 \times 10^{-3}\,\mathrm{arcsec}^2$ & $3.1 \times 10^{-8}\,\mathrm{arcsec}^4$ & $2.0 \times 10^{-3}$ & $3.0 \times 10^{-5}$ \\
    LSST & $1.5 \times 10^{-3}\,\mathrm{arcsec}^2$ & $1.8 \times 10^{-7}\,\mathrm{arcsec}^4$ & $2.7 \times 10^{-4}\,\mathrm{arcsec}^2$ & $2.9 \times 10^{-9}\,\mathrm{arcsec}^4$ & $5.5 \times 10^{-4}$ & $4.9 \times 10^{-6}$
  \enddata
  \tablecomments{Requirements on bias parameters for differential chromatic refraction ($\Delta \bar{R}$ and $\Delta V$) and for chromatic seeing ($\Delta r^2_\mathrm{PSF}/r^2_\mathrm{PSF}$), defined in Equations~\ref{eqn:Rbar}, \ref{eqn:V}, and \ref{eqn:delta_r2}.
  The means and variances are evaluated for all star-galaxy SED pairs within each tomographic redshift bin.
  The values of chromatic bias means are those for which each bias by itself would degrade the accuracy and/or precision of the survey by an amount equivalent to the statistical sensitivity.
  The values of chromatic bias variances are those for which each bias by itself could (but will not necessarily) degrade the accuracy and/or precision of the survey by an amount equivalent to the statistical sensitivity.}
\end{deluxetable*}

We propagate the requirements on $|\langle m\rangle|_\mathrm{max}$ and $\mathrm{Var}(c)_\mathrm{max}$ to requirements on the mean and variance of $(\Delta \bar{R}_{45})^2$ using the typical values of $m/(\Delta \bar{R}_{45})^2 \approx -2\,\mathrm{arcsec}^{-2}$ and $c/(\Delta \bar{R}_{45})^2 \approx 1\,\mathrm{arcsec}^{-2}$ plotted in Figure \ref{fig:misregistration_bias}.
Similarly, we use Equations \ref{eqn:m1m2_DCR}-\ref{eqn:c2_DCR} and \ref{eqn:m1m2_seeing}-\ref{eqn:c2_seeing} to set requirements on the mean and variance of $\Delta V$ and $\Delta r^2_\mathrm{PSF}/r^2_\mathrm{PSF}$.
We stress that the requirements $|\langle m\rangle|_\mathrm{max}$ describe the point at which the {\it combined} uncertainties from {\it all} systematic effects will equal the expected statistical uncertainties of weak lensing surveys, and thus, our ultimate goal will be to correct chromatic biases to well below these values.
Similarly, the requirements $\mathrm{Var}(c)_\mathrm{max}$ describe an upper limit to the point at which combined systematic uncertainties will equal the expected statistical uncertainties.

The final ingredients needed to convert requirements on $m$ and $c$ into requirements on the chromatic bias parameters $(\Delta \bar{R}_{45})^2$, $\Delta V$, and $\Delta r^2_\mathrm{PSF}/r^2_\mathrm{PSF}$ are the typical galaxy sizes $r^2_\mathrm{gal}$ and typical PSF sizes $r^2_\mathrm{PSF}$, which we summarize in Table~\ref{table:sizes} and describe below.
The typical PSF ellipticity $\chi_\mathrm{PSF}$ is also needed to set a requirement on $\mathrm{Var}(\Delta r^2_\mathrm{PSF}/r^2_\mathrm{PSF})$, which arises due to the requirement on $\mathrm{Var}(c)$ and the additive biases due to chromatic seeing (Equations \ref{eqn:c1_seeing} and \ref{eqn:c2_seeing}).
We will conservatively set $\chi^\mathrm{PSF}=0.05$ \citep{Jee+Tyson11}.

As for the galaxy and PSF sizes, we can generically assume that these will be of the same order, as surveys will naturally attempt to measure the shapes of galaxies with sizes down to the survey resolution limits.
For the PSF size, we use estimates of the typical PSF FWHM of $0.7$ arcseconds for LSST~\citep{LSSTSB} and $0.8$ arcseconds for DES~\citep{DES++07}\footnote{In this study, we do {\em not} degrade these PSF values for the increased air mass at non-zero zenith angles.}, and note that a Moffat profile with parameter $\beta=3.0$ (which is typical for an atmospheric PSF) has $\sqrt{r^2}/\mathrm{FWHM} \approx 1$.
For the galaxy sizes, we use the LSST simulated galaxy catalog \textsc{CatSim}~\citep{Connolly++inprep} (described in Section \ref{sec:catalog}) and
apply a magnitude limit of $i_\mathrm{AB} < 25.3$ and 24.0 for LSST and DES, respectively.
As a measure of typical galaxy size, we use the mode of the distribution of $\sqrt{r^2_\mathrm{gal}}$ for galaxies near the peak of the DES and LSST source-galaxy redshift distributions.
We find a typical galaxy size corresponding to $r^2_\mathrm{gal} \approx (0.3\arcsec)^2$ and $(0.4\arcsec)^2$ for LSST and DES, respectively.

Folding everything together, we summarize in Table~\ref{table:chromatic_requirements} the maximum allowed values of the LSST and DES chromatic bias parameters --
$\langle(\Delta \bar{R}_{45})^2\rangle$, $\mathrm{Var}((\Delta \bar{R}_{45})^2)$, $|\langle\Delta V\rangle|$, and $\mathrm{Var}(\Delta V)$ for DCR, and $|\langle\Delta r^2_\mathrm{PSF}/r^2_\mathrm{PSF}\rangle|$ and $\mathrm{Var}(\Delta r^2_\mathrm{PSF}/r^2_\mathrm{PSF})$ for chromatic seeing --
for which each bias by itself would degrade the constraining power of the survey by an amount equivalent to the statistical sensitivity.

\section{Chromatic biases in simulated catalogs} %4
\label{sec:catalog}
To estimate the sizes of chromatic biases in a real survey, we require a realistic catalog of stars and galaxies containing an accurate distribution of SEDs.
We use simulated star and galaxy catalogs generated by the LSST catalog simulator \textsc{CatSim} \citep{Connolly++inprep}.
While we use parameters most appropriate for LSST, the results are also largely applicable to other ground-based surveys such as DES.

The \textsc{CatSim} stellar SEDs are described by \citet{Kurucz93} models, and are distributed in space and metallicity according to the Milky Way model in \citet{Juric++08}, \citet{Ivezic++08}, and \citet{Bond++10}.
For stars, we restrict ourselves to objects with $17 < i_\mathrm{AB} < 22$, which yields stars that are faint enough so they do not saturate the CCD in a 15-second LSST exposure, yet bright enough to provide reasonable constraints on the PSF in a single exposure.

The \textsc{CatSim} galaxy SEDs are generated starting with the simulated galaxy catalog of \citet{DeLucia++06}, in which each galaxy is characterized by its redshift and a set of parameters for the bulge and disk component of the galaxy: color ($B$-, $V$-, $R$-, $I$-, and $K$-band magnitudes), size, dust estimates, and stellar population age estimates.
A sophisticated fitting program then finds the best fit SED parameters and galaxy extinction that reproduce the color information for each galaxy.
The bulge and disk SEDs are parameterized with the single stellar population models of \citet{Bruzual+Charlot03}.
We restrict ourselves to objects with $i_\mathrm{AB} < 25.3$, the so-called LSST `gold sample', though the base \textsc{CatSim} catalog goes several magnitudes deeper.

\begin{figure*}
  \begin{center}
    \epsscale{1.15}
    \plotone{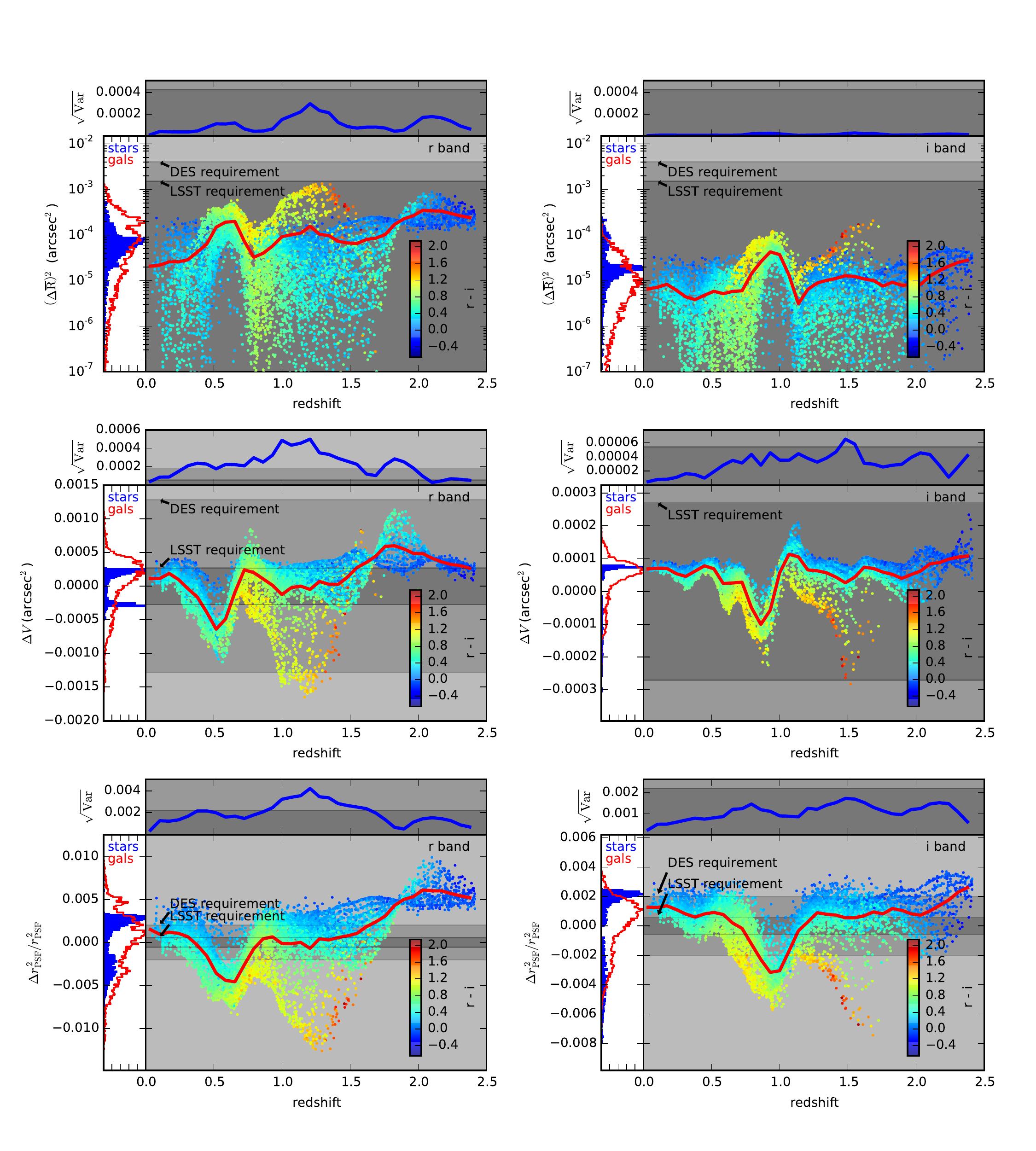}
  \end{center}
  \caption{
    \label{fig:bias_panel}
      Chromatic biases.
      \emph{Top:} Squared relative shifts in PSF centroid, $(\Delta \bar{R})^2$, due to DCR for objects with different SEDs, in LSST $r$-band (left) and $i$-band (right) for a zenith angle of 45 degrees.
      \emph{Middle:} Relative shifts in second central moment, $\Delta V$, due to DCR for objects with different SEDs, in LSST $r$-band (left) and $i$-band (right) for a zenith angle of 45 degrees.
      \emph{Bottom:} Relative fractional shifts in second-moment squared radius, $\Delta r^2_\mathrm{PSF}/r^2_\mathrm{PSF}$, due to chromatic seeing for objects with different SEDs, in LSST $r$-band (left) and $i$-band (right).
      In each panel, the shifts are calculated relative to the mean stellar shift: $\Delta X = X - \langle X \rangle_\mathrm{stars}$.
    The blue histograms on the left axes correspond to 4\,000 stars with $17 < i_\mathrm{AB} < 22$ in the LSST simulated catalog \textsc{CatSim}.
    The red projected histograms correspond to 4\,000 galaxies with $i_\mathrm{AB} < 25.3$ in \textsc{CatSim}.
    Each galaxy is represented by a colored dot and plotted as a function of redshift, with the color scale (shown on the plot) indicating the observer-frame $r - i$ color of each galaxy.
    The red (blue) curves indicate the running means (square-root-variances) of the relative galactic shift.
    The gray bands in the lower panels indicate requirements on the mean shifts such that the resulting systematic uncertainty is less than the expected statistical uncertainty for the entire survey for DES (light gray outer band) and LSST (dark gray inner band).
    The shaded bands in the upper panels indicate the sufficient (but not necessary) requirements on the square-root-variance.
    Note that the plotting range is set to highlight the data and thus each requirement band may not be entirely visible.
  }
\end{figure*}

In Figure~\ref{fig:bias_panel}, we plot the distributions of chromatic biases due to DCR for a zenith angle of 45 degrees (squared centroid shift $(\Delta \bar{R})^2$ and zenith-direction variance shift $\Delta V$), and chromatic seeing (fractional shift in second-moment squared radius, $\Delta r^2_\mathrm{PSF}/r^2_\mathrm{PSF}$) for 4\,000 galaxies and 4\,000 stars from the restricted \textsc{CatSim} catalogs.
In each case, the chromatic biases are calculated relative to the mean bias of the stars.
We find that the mean star, in terms of chromatic biases, has approximately a K5V SED.
The distribution for stars is shown as the blue histogram to the left of each scatter plot.
The distribution for galaxies is shown as a function of redshift in the scatter plot and projected across redshift as the red histogram to the left.
The colors of the plotting symbols indicate the $r - i$ color of each galaxy in the observer's frame of reference.
As expected, for bluer spectra, the PSFs are generally stretched more along the zenith direction, and have larger fractional shifts in second-moment squared radius.

Each figure also shows a thick red line to indicate the running mean of the galactic biases.
Since the biases are given relative to the mean stellar bias, this line indicates the running mean differential bias between all star-galaxy pairs.
This is precisely the meaning of the quantities in angle brackets, $\langle \cdot\cdot\cdot \rangle$, as used in Table~\ref{table:chromatic_requirements};
therefore, these are three of the quantities that must be corrected if they are not small compared to the values of $\langle(\Delta \bar{R}_{45})^2\rangle$, $|\langle\Delta V\rangle|$, and $|\langle\Delta r^2_\mathrm{PSF}/r^2_\mathrm{PSF}\rangle|$ in Table\,\ref{table:chromatic_requirements}.
The dark gray inner bands and light gray outer bands in the lower panels of each figure span the range $\pm |\langle \cdot\cdot\cdot \rangle|$ for LSST and DES, respectively, for each mean differential bias.
The shaded bands in the upper panels indicate the sufficient (but not necessary) requirements on the square-root-variance.
Hence, in total these three figures (in six panels) encode 24 different requirements; each panel displays both a multiplicative (running mean) and an additive (running variance) requirement for each of two experiments.
Finally, we note that the full extent of the DES requirement band is not visible in some panels.

\begin{deluxetable}{lcccc}
  \tablecaption{
    \label{table:biases_summary}
    Impact of chromatic biases.
  }
  \tablehead{
    &
    \multicolumn{2}{c}{DES} &
    \multicolumn{2}{c}{LSST} \\
    \colhead{Chromatic bias} &
    \colhead{$r$-band} &
    \colhead{$i$-band} &
    \colhead{$r$-band} &
    \colhead{$i$-band}
    }
  \startdata
% bias                                                    DES-r    DES-i    LSST-r   LSST-i
  {\it Multiplicative biases}\phantom{\Large I} & \multicolumn{4}{l}{Necessary conditions}\\
  $\langle(\Delta\bar{R}_{45})^2\rangle$                   & \happy (\happy)  & \happy (\happy)  & \happy (\happy)  & \happy (\happy) \\
  $\langle\Delta V\rangle$                                & \bored (\happy)  & \happy (\happy)  & \sad (\bored)  & \bored (\happy) \\
  $\langle \Delta r^2_\mathrm{PSF}/r^2_\mathrm{PSF}\rangle$ & \angry (\sad)  & \sad (\bored)  & \angry (\sad)  & \angry (\bored) \\
  {{\it Additive biases}\tablenotemark{*}}\phantom{\huge I} & \multicolumn{4}{l}{Sufficient but not necessary conditions}\\
  $\Var((\Delta\bar{R}_{45})^2)$                           & \happy (\happy)  & \happy (\happy)  & \bored (\happy)  & \happy (\happy) \\
  $\Var(\Delta V)$                                        & \angry (\angry)  & \bored (\happy)  & \angry (\angry)  & \sad (\sad) \\
  $\Var(\Delta r^2_\mathrm{PSF}/r^2_\mathrm{PSF})$          & \bored (\happy)  & \happy (\happy)  & \sad (\bored)  & \bored (\bored)
  \enddata
  \tablecomments{
    Qualitative assessment of the impact of different chromatic effects for DES and LSST in $r$- and $i$-band.
    The impact of each chromatic effect is ranked on a scale \happy $\rightarrow$ \bored $\rightarrow$ \sad $\rightarrow$ \angry, where \happy indicates the estimated effect size is well below the requirements derived in Section \ref{sec:requirements}, \bored indicates the bias is a significant fraction of the requirement, \sad indicates that the requirement is exceeded, and \angry indicates the requirement is greatly exceeded.
    Faces in parentheses indicate the size of the residual effects after the machine-learning corrections described in Section \ref{sec:machine_learning} are applied.
    Recall that requirements are set such that the systematic uncertainty incurred for a given effect and experiment is equal to that experiment's statistical uncertainty.
  }
  \tablenotetext{*}{Recall that requirements on variances are sufficient but not necessary to keep the associated systematic uncertainty at or below the statistical uncertainty, and should hence be viewed as pessimistic.}
\end{deluxetable}

Briefly, Figure \ref{fig:bias_panel} indicates that effects of centroid shifts due to DCR (top panels) are generally well below the threshold for concern for either experiment in either band.
Second-moment shifts due to DCR (middle panels) have somewhat larger effects, especially in $r$-band and especially for LSST.
The mean shift in relative PSF size due to chromatic seeing (bottom panels) affects both experiments significantly, though the variance of the relative PSF size difference only affects LSST significantly.
Table \ref{table:biases_summary} provides a summary of the qualitative impacts of DCR and chromatic seeing for DES and LSST in $r$- and $i$-band.

Finally, we note that additional chromatic effects, such as those originating in the telescope optics and camera sensors, will also impart SED-dependent PSF biases on galaxy shape measurements (see Section~\ref{sec:other_chroma}).

\section{Predicting chromatic biases from photometry} %5
\label{sec:machine_learning}
\begin{figure*}
  \begin{center}
    \epsscale{1.15}
    \plotone{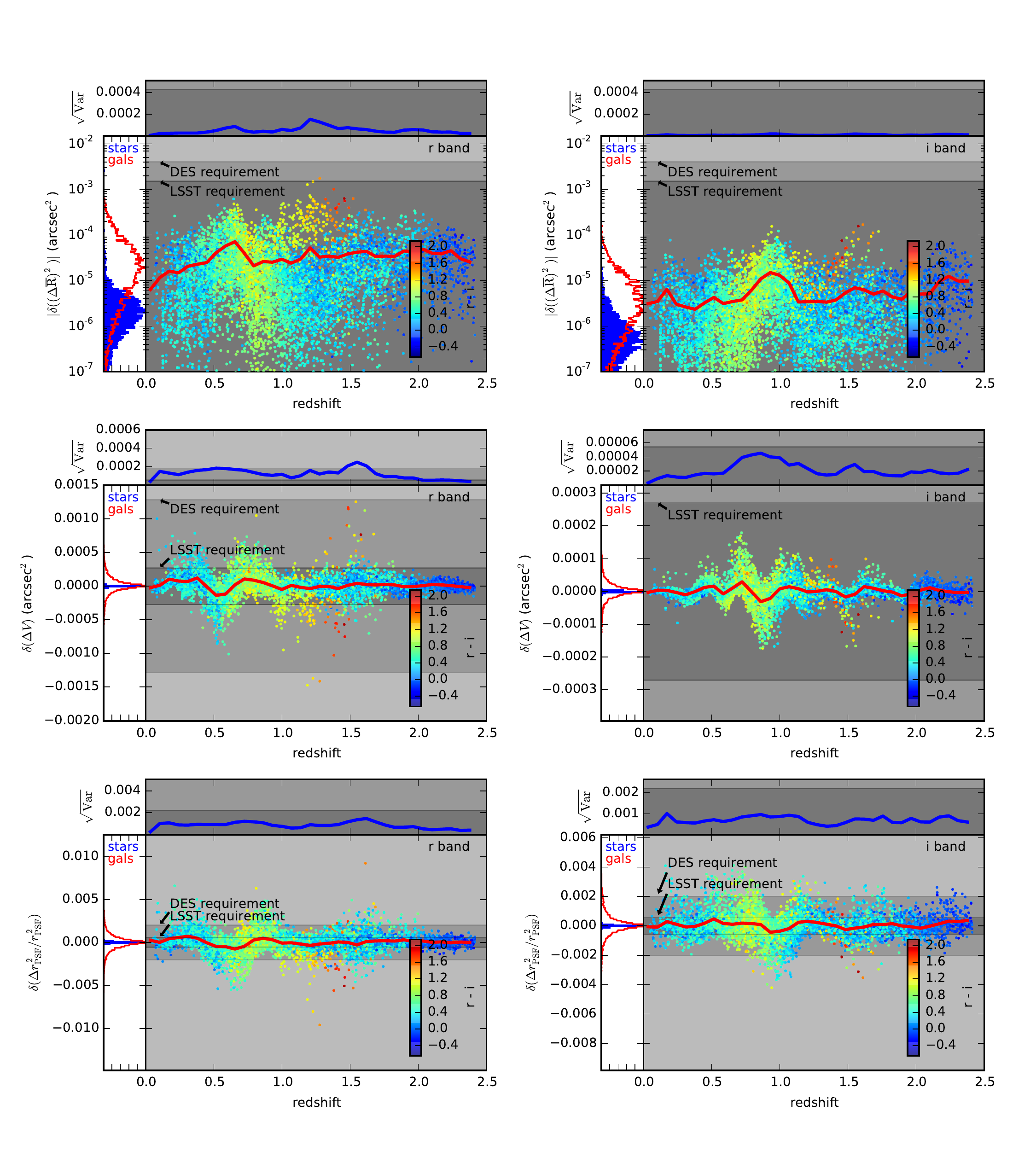}
  \end{center}
  \caption{
    \label{fig:corrected_bias_panel}
    Chromatic bias residuals after applying Extra Trees Regression to estimate the biases from six-band photometry.
    \emph{Top:} Squared PSF centroid shift due to DCR.
    \emph{Middle:} PSF zenith-direction second-moment shift due to DCR.
    \emph{Bottom:} Fractional second-moment squared radius shift due to chromatic seeing.
    See the caption for Fig.~\ref{fig:bias_panel} for a description of the histograms, colored points and curves, and gray bands.
  }
\end{figure*}

As can be seen by eye in Figure~\ref{fig:bias_panel}, the chromatic biases $(\Delta \bar{R})^2$,  $\Delta V$, and $\Delta r^2_\mathrm{PSF}/r^2_\mathrm{PSF}$ are correlated with the $r - i$ color of the SED (indicated by the color of the dot representing each galaxy).
This is expected since the magnitude of each chromatic effect depends directly on the shape of the SED across a single filter (see Equations~\ref{eqn:Rbar}, \ref{eqn:V}, and \ref{eqn:seeing_M2_SED}) and the shape within a particular filter is correlated with the flux difference between neighboring filters.
Exploiting this correlation was investigated by PB12 as one method that potentially could be used to correct for the differences in the effective PSFs of stars and galaxies when chromatic effects are important.
PB12 concluded that such an approach would be insufficient for LSST, especially in bluer filters.
Here we investigate a natural extension to the single-color correction, which is to estimate a correction using photometric information from all available filters.

\subsection{Extra Trees Regression}
\label{sec:ETR}
The task of predicting a parameter such as the centroid shift $\Delta \bar{R}$ from features such as $ugrizy$ photometry falls into the category of a regression problem in supervised machine learning -- and is very similar to the problem of estimating redshifts from photometry.
A wide variety of supervised-learning algorithms have been used to estimate photometric redshifts \citep{Zheng+Zhang12}.
Here we apply the Extra Trees Regression (ETR) algorithm~\citep{Geurtz++06}, using the implementation in the Python \textsc{scikit-learn} package \citep{Pedregosa++11}.
(We also investigated other machine learning algorithms, including Support Vector Regression and Random Forest Regression. ETR performed better than the others.)

Separately for both the star and galaxy simulated LSST catalogs described in the previous section, we designate members of a training set (16\,000 objects) and a validation set (4\,000 objects).
The stellar and galactic validation-set SEDs are identical to those used to create Figure~\ref{fig:bias_panel}, and disjoint from the training-set SEDs.
The input data for each object in the training and validation sets consist of five colors ($u-g$, $g-r$, $r-i$, $i-z$, $z-y$) and the $i$-band magnitude, calculated by integrating the catalog SEDs over each of the six LSST filters.
The input for the training set additionally includes the true chromatic bias parameters $\Delta \bar{R}$,  $\Delta V$, and $\Delta r^2_\mathrm{PSF}/r^2_\mathrm{PSF}$, calculated via Equations~\ref{eqn:Rbar}, \ref{eqn:V}, and \ref{eqn:delta_r2}.
The training set is used to determine the ensemble of decision trees used by ETR that are then used to predict the chromatic bias parameters from the validation-set photometry.
To incorporate observational uncertainties, the magnitude in each band for each object in the validation set is perturbed by a Gaussian with width equal to the expected photometric uncertainty at the end of the 10-year LSST survey (see \ref{sec:magerr}).
We set the minimum photometric uncertainty for each band to 0.01 magnitudes.
Note that the additional photometric uncertainty for DES is not estimated.

In Figure \ref{fig:corrected_bias_panel}, we show the residuals of the chromatic bias parameters calculated as the difference between the true bias ($\Delta (\bar{R})^2$,  $\Delta V$, or $\Delta r^2_\mathrm{PSF}/r^2_\mathrm{PSF}$) and the ETR prediction, for each star and galaxy in the validation set:
\begin{equation}
  \label{eqn:deltaDeltaRSqr}
  \delta((\Delta \bar{R})^2) = 2 (\Delta \bar R) \delta(\Delta \bar{R}),
\end{equation}

\begin{equation}
  \label{eqn:deltaDeltaR}
  \delta(\Delta \bar{R}) = \Delta \bar{R} - \Delta \bar{R}_{\rm ETR},
\end{equation}

\begin{equation}
  \label{eqn:deltaDeltaV}
  \delta(\Delta \bar{V}) = \Delta \bar{V} - \Delta \bar{V}_{\rm ETR},\quad{\rm and}
\end{equation}

\begin{equation}
  \label{eqn:deltaDeltar2}
  \delta(\Delta r^2_\mathrm{PSF}/r^2_\mathrm{PSF}) = \Delta r^2_\mathrm{PSF}/r^2_\mathrm{PSF} - (\Delta r^2_\mathrm{PSF}/r^2_\mathrm{PSF})_\mathrm{ETR}.
\end{equation}

Note that the ranges on the vertical axes in these plots of residuals are the same as the ranges used in the corresponding plots showing the raw biases (Figure~\ref{fig:bias_panel}).
In all cases, ETR significantly reduces both the mean and the variance of the chromatic bias.
The most significant remaining biases are the $r$-band variance of second-moment shifts due to DCR (middle left panel), and the mean relative PSF size difference due to chromatic seeing (bottom panels).
We will address further mitigation of these biases in Section \ref{sec:psf_corr_calibrate}.

\subsection{Additive biases}
\label{sec:additive_biases}

In Section \ref{sec:requirements}, we noted that uncorrelated additive shear biases can only affect shear power spectra at $\ell=0$, and hence do not affect cosmological constraints derived from shear power spectra.
On the other hand, scale-dependent bias correlations \emph{do} impact cosmological constraints.
For atmospheric chromatic biases, the most likely source of scale-dependent correlations originates with the color dependence of galaxy clustering \citep{Balogh++99} -- red galaxies are more tightly clustered than blue galaxies.
This correlation, together with the color dependence of atmospheric chromatic biases, leads to scale-dependent chromatic biases.
While a full treatment of this effect requires a high-fidelity catalog of galaxy SEDs as a function of right ascension, declination, and redshift, which is beyond the scope of this study, we demonstrate that angular correlations between the {\it residual} chromatic biases are likely to be significantly mitigated by the machine-learning corrections.

\begin{figure*}
  \begin{center}
    \epsscale{1.15}
    \plotone{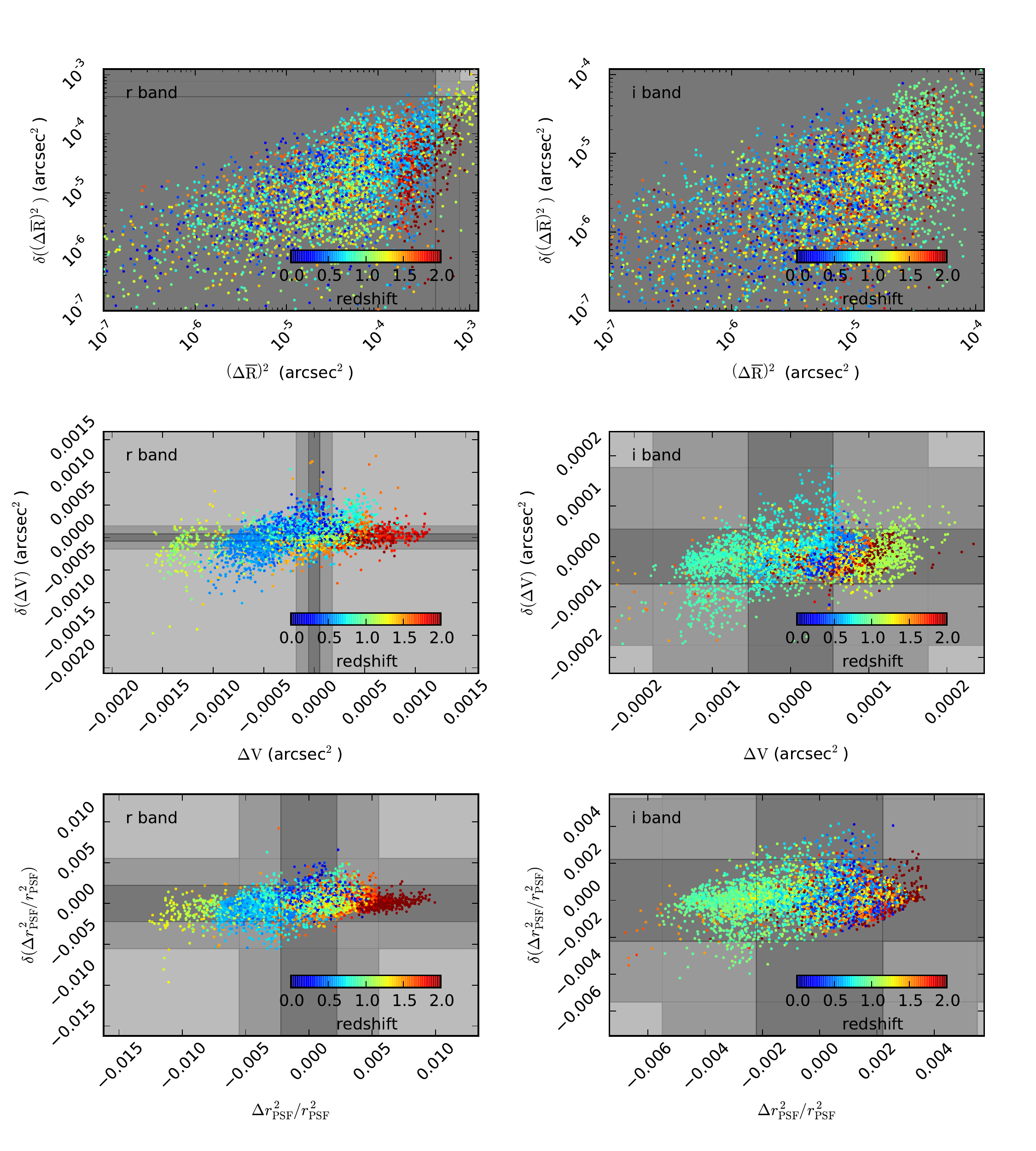}
  \end{center}
  \caption{
    \label{fig:bias_vs_corrected}
    Chromatic PSF bias residuals versus uncorrected values for SEDs
    from the LSST \textsc{CatSim} galaxy catalog.  Each SED is
    represented by a colored point, with the color scale indicating
    the redshift.  The shaded bands indicate requirements on each
    chromatic bias \emph{square-root-variance} such that the resulting
    systematic uncertainty is less than the expected statistical
    uncertainty for the entire survey for DES (light gray outer band)
    and LSST (dark gray inner band).  Note that the bands sometimes
    extend beyond the plotted regions.
    }
\end{figure*}

In Figure \ref{fig:bias_vs_corrected}, we plot the residual chromatic biases $\delta((\Delta \bar{R})^2)$, $\delta(\Delta V)$, and $\delta(\Delta r^2_\mathrm{PSF}/r^2_\mathrm{PSF})$ versus their uncorrected values, with the color of each point indicating the redshift of the galaxy SED.
One can immediately see that the variance of each bias has been decreased, and more importantly that the residuals are not strongly correlated with the uncorrected biases.
This lack of correlation implies that external variables initially correlated with chromatic biases, such as galaxy clustering, are likely significantly less correlated with the residuals.

Since the requirements we derived for chromatic bias (residual) variances assume maximally spatially correlated additive biases, and we predict minimal correlations in practice, it is unlikely that any of these residual variances (for instance as present for $r$-band $\Delta V$) will lead to systematic uncertainties larger than DES or LSST statistical uncertainties.

\subsection{Limitations of Analytic Estimates and Machine Learning Predictions of Chromatic Biases}
\label{sec:limitations}

The previous sections indicate that chromatic biases can be predicted from six-band photometry with enough accuracy to prevent chromatic effects from dominating the LSST cosmic shear error budget.
However, these conclusions depend on several assumptions that we make explicit here.

\begin{enumerate}

\item The analytic calculations in Section\,\ref{sec:analytics} are based on the additivity of second moments of the galaxy profile and the PSF under convolution (Equation~\ref{eqn:second_moments_add}), which is generally the case only for unweighted second moments.
However, all practical shape measurement algorithms require weighted second moments, including model-fitting algorithms in which the model being fit is equivalent to a weight function.
As we will show in the next section, analytic predictions begin to break down for practical algorithms.

\item The accuracy of chromatic bias corrections predicted from six-band photometry depends on the fidelity of the SEDs for both stars and galaxies in the catalogs used to train the machine-learning algorithm.

\item Our treatment of the angular covariance of chromatic biases is limited by the lack of realistic clustering properties of galaxies with different SEDs in the \textsc{CatSim} catalog.

\item We have calculated biases and requirements assuming typical values for PSF and galaxy sizes.
  Chromatic biases will increase for larger PSFs or smaller galaxies, and decrease for smaller PSFs or larger galaxies.

\item We have assumed that the PSF at one wavelength is simply a scaled and shifted version of the PSF at any other wavelength.
  This is likely to be the case on average, but may not be exactly true for single instances of the PSF.

\item We have focused here only on chromatic PSF effects originating in the atmosphere.
  While we expect atmospheric chromatic effects to dominate, additional chromatic effects will also arise from optics and sensors (see Section~\ref{sec:other_chroma}).
  These effects may complicate the above analysis and the following correction techniques.

\end{enumerate}

\section{PSF model-fitting bias} %6
\label{sec:model_fitting}
The analytic predictions for chromatic biases described in the previous section depend on the second-moment squared radii of the PSF and galaxy profile -- but not on the exact shape of their profiles.
In this section, we describe an investigation (based on a `ring test') of how chromatic biases depend on the detailed shape of the PSF and galaxy profiles, and on different measures of `size' -- second-moment squared radius ($r^2$), full width at half maximum (${\rm FWHM}$), or half-light radius (${\rm HLR}$).
We show that, in addition to the chromatic biases predicted by analytic equations, there exists a `model bias'.
If the apparent galaxy image is deconvolved with the stellar PSF, the chromatic bias is not necessarily the same as the analytic prediction and therefore cannot be completely accurately corrected at the `catalog level' -- for example, with a machine-learning algorithm based only on photometry.
We begin this section with a review of the ring test.

\subsection{Ring Test}\label{sec:ringtest}

An alternative way to estimate the shear measurement bias induced by chromatic effects is to generate simulated galaxy images using the effective PSF of the galaxy and then attempt to recover the reduced shear assuming that the effective PSF is that which would have been measured from a star.
A ring test \citep{Nakajima+Bernstein07} is a specific prescription for generating a suite of such simulations, designed to rapidly converge to the precise (though biased by the use of the wrong PSF) value of the mean reduced-shear estimator $\boldsymbol{\hat{g}}$ for a given true input shear $\boldsymbol{g}$.
The name `ring test' is derived from the arrangement of intrinsic galaxy shape parameters $(\epsilon^{(i)}_1, \epsilon^{(i)}_2)$ used in the simulated images, which form a ring in complex ellipticity space centered at the origin (i.e., $|\boldsymbol{\epsilon}^{(i)}|$ is constant) before shearing.
By rotating the simulated galaxy in real space such that the intrinsic ellipticities exactly average to zero for each pair of galaxies (i.e., the complex ellipticities lie on opposite sides of a ring),
the results of the test converge faster than for randomly (though isotropically) chosen intrinsic ellipticities, which only average to zero statistically.
The ring test is best done using $\epsilon$-ellipticities, as these average together to precisely yield the applied reduced shear $g$.
In contrast, $\chi$-ellipticities only approximately yield $2 g$.

For a parametrically defined galaxy profile, where the apparent ellipticity $\boldsymbol{\epsilon^{(a)}}$ contributes two real parameters, the ring test can be implemented as follows.
\begin{enumerate}
    \item Choose a PSF shape and generate ${\rm PSF}_{\rm eff}^*$ and ${\rm PSF}_{\rm eff}^{\rm g}$ according to Equation~\ref{eqn:PSFeff} for a stellar and galactic SED, respectively.
    \item Choose a circularly symmetric fiducial galaxy profile -- e.g., a Sérsic profile\footnote{A circular Sérsic profile has functional form $I(r) \propto e^{-k (r/r_e)^{1/n}}$.  The Sérsic index $n$ sets the sharpness of the central peak and the importance of the profile wings.  Gaussian, exponential, and de~Vaucouleurs profiles are recovered when $n$ equals 0.5, 1.0, and 4.0, respectively.} with half-light radius $r_e$ and Sérsic index $n$.
    \item Choose an ellipticity magnitude $\epsilon^{(i)}$ for the intrinsic galaxy profile.
    We use $\epsilon^{(i)}=0.3$ for all the investigations presented here, though one could also draw $\epsilon^{(i)}$ from an intrinsic ellipticity distribution in this step.
    \item Choose a fiducial reduced shear $\boldsymbol{g}$.
    We use $\boldsymbol{g}=(0.0, 0.0)$ and $\boldsymbol{g}=(0.01, 0.02)$ for all the investigations presented here.
    \item Choose an intrinsic ellipticity $\boldsymbol{\epsilon^{(i)}}$ on the ellipticity ring with magnitude $\epsilon^{(i)}$ determined above (i.e., choose an angle in complex ellipticity space).
    \item Apply this ellipticity to the circular galaxy profile.
    \item Apply shear to the galaxy; the resulting apparent ellipticity $\boldsymbol{\epsilon^{(a)}}$ is given by Equation~\ref{eqn:applyshear_eps}.
    \item Generate a target image by convolving the galaxy profile with the galactic PSF (${\rm PSF}_{\rm eff}^{\rm g}$) and spatially integrating over each pixel.
    \item Using a stellar PSF (${\rm PSF}_{\rm eff}^*$), estimate the ellipticity of the target image.
        This could be done with any number of shape measurement algorithms, but for the present study, we simply minimize (over the galaxy model parameters) the sum (over pixels) of the squared differences between the target image and the image formed by convolving the galaxy model with the stellar PSF.
        The ellipticity estimate is just the value of the ellipticity parameter that minimizes this statistic.
    \item Repeat steps 6-9 using the opposite intrinsic ellipticity: $\boldsymbol{\epsilon^{(i)}} \rightarrow -\boldsymbol{\epsilon^{(i)}}$.
    \item Repeat steps 5-10 for as many values around the ellipticity ring as desired.
          We have found that the minimum number of pairs uniformly spaced around the ellipticity ring required for convergence of the test is three, and that using more pairs than this does not change our results.
    \item Average all recorded ellipticity estimates.
          This is the shear estimator $\boldsymbol{\hat{g}}$.
    \item Repeat steps 3-12 to map out the relation $\boldsymbol{g} \rightarrow \boldsymbol{\hat{g}}$.
    \item From Equation~\ref{eqn:shearbias}, $1+m_i$ and $c_i$ are then the slope and intercept of the best-fit linear relation between $g_i$ and $\hat{g_i}$.
    \item $m_i$ and $c_i$ may depend on parameters such as the Sérsic index of the galaxy, the Moffat profile index $\beta$, and so on.
    These dependencies can be investigated by repeating steps 1-14.
\end{enumerate}

Here we investigate using a single Sérsic profile as the galaxy model.
The Sérsic profile has seven parameters: the $x$ and $y$ coordinates of the center, the total flux, the half light radius $r_e$ (also called the effective radius), the
two-component apparent ellipticity $\boldsymbol{e^{(a)}}$, and the Sérsic index $n$.

We start with a monochromatic PSF that has either a Gaussian or $\beta=3.0$ Moffat profile.
To make the PSF chromatic, we allow the centroid and the size to vary with wavelength.
Schematically,
\begin{equation}
  \label{eqn:chromatic_PSF}
  \mathrm{PSF}(x, y, \lambda) = f\left(x, y-y_0(\lambda), {s}(\lambda)\right),
\end{equation}
where $f$ is either a normalized circular Gaussian or normalized circular Moffat profile with arguments for the centroid and size,
\begin{equation}
  y_0(\lambda) = R(\lambda) - R_0,
\end{equation}
\begin{equation}
  {s}(\lambda) = {s}_0 \left(\frac{\lambda}{\lambda_0}\right)^{-1/5},
\end{equation}
and $R_0$ and $s_0$ are the refraction (due to DCR) and size (due to seeing) -- either FWHM or $\sqrt{r^2}$ -- of the PSF at the fiducial wavelength $\lambda_0$
The effective PSF for the star or galaxy is then given by the (normalized) integral over wavelength of the product of the PSF in Equation~\ref{eqn:chromatic_PSF}, the SED, and the transmission function, as described in Equation~\ref{eqn:PSFeff}.
The results presented here all assume that the direction of refraction (the parallactic angle) is aligned with the simulated pixel grid, though we have checked that changing this direction does not affect our results.

\subsection{Dependence of Predictions for Chromatic Effects on
Shapes of PSF and Galaxy Profile}\label{sec:nongaussian}

To investigate the dependence of chromatic effects on the detailed shapes of the PSF and the galaxy profile, we first choose fiducial profiles about which to vary the shapes while keeping measures of size constant.
Our fiducial PSF and galaxy profiles are both Gaussian.
For the galaxy, we set the squared second-moment radius to $(0.3\arcsec)^2$, which is the typical size expected for LSST source galaxies (see Section~\ref{sec:requirements}), and set the intrinsic ellipticity $|\boldsymbol{\epsilon}^{(i)}|$ to $0.3$.
For the PSF, we set the FWHM to $0.7\arcsec$ at the effective wavelength of the filter, and ellipticity at any given wavelength to 0.0 (though DCR will make the ellipticity of the effective PSF non-zero).
The variations from our fiducial model include changing the galaxy to a de~Vaucouleurs profile, changing the PSF to a Moffat ($\beta = 3$) profile, and changing both galaxy and PSF profiles simultaneously.
Throughout this section and the next, we assume an LSST $r$-band filter and a zenith angle of 45 degrees, but remind the reader that approximately 80\% of LSST observations will occur at smaller zenith angles.
To investigate particularly pernicious chromatic biases, we choose a relatively red SED for the galaxy (an Sa template), and a relatively blue SED for the star (a G5V template).
We simulate pixels that are $0.2\arcsec$ on a side and an image that is 31 pixels on a side; we have checked that our results are insensitive to moderate variations in these parameters.

\begin{figure}
  \begin{center}
    \epsscale{1.2}
    \plotone{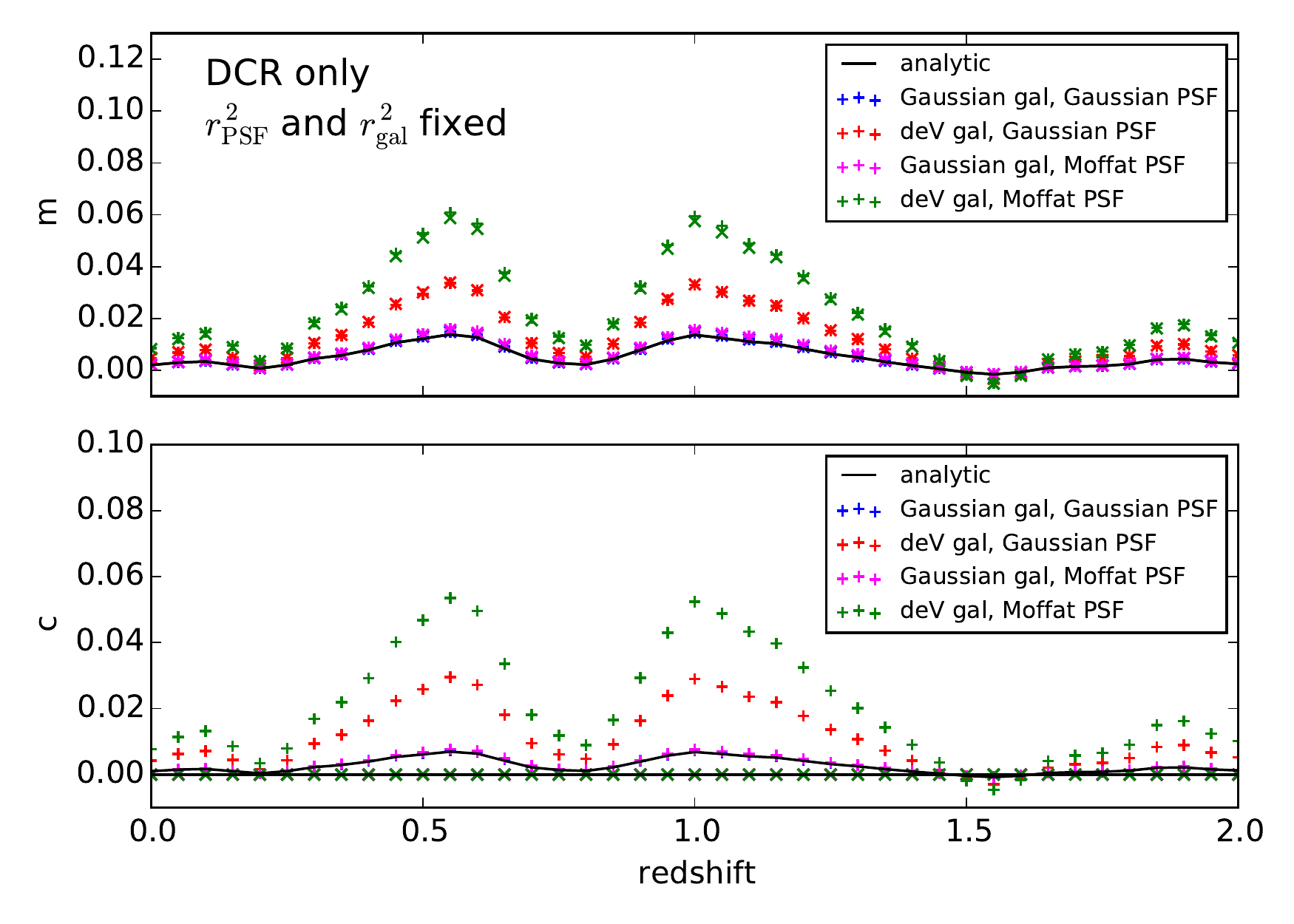}
  \end{center}
  \caption{
    \label{fig:r2_DCR_modelfit}
    The shear calibration parameters due to differential chromatic refraction as a function of galaxy redshift, predicted analytically (solid black curve) and with a
    ring-test (colored symbols).
    Chromatic seeing is not implemented here.
    The filter is the LSST $r$-band, and the zenith angle is 45 degrees.
    The predictions are based on a G5V stellar SED and an Sa galactic SED.
    The galaxy has intrinsic ellipticity $|\boldsymbol{\epsilon}^{(i)}| = 0.3$ and second-moment squared radius $r^2_\mathrm{gal} = (0.3\arcsec)^2$.
    The PSF has second-moment squared radius $r^2_\mathrm{PSF} = (0.42\arcsec)^2$ measured at the filter effective wavelength.
    The analytic prediction (solid curves) depends on the galaxy second-moment squared radius, but is otherwise independent of the profiles of the PSF and galaxy.
    $+$'s indicate the components of $m$ and $c$ oriented along and perpendicular to the zenith direction, and $\times$'s indicate the components oriented 45 degrees to the zenith direction.
    The different colored plotting symbols indicate different assumed profiles for the galaxy and for the fixed-wavelength PSF, with $r^2$ fixed for both the galaxy and the PSF in all cases.
    The blue symbols lie directly under the magenta symbols (and are therefore not visible) since the ring-test predictions do not depend on the PSF profile when $r^2_\mathrm{PSF}$ is fixed.
  }
\end{figure}

\begin{figure}
  \begin{center}
    \epsscale{1.2}
    \plotone{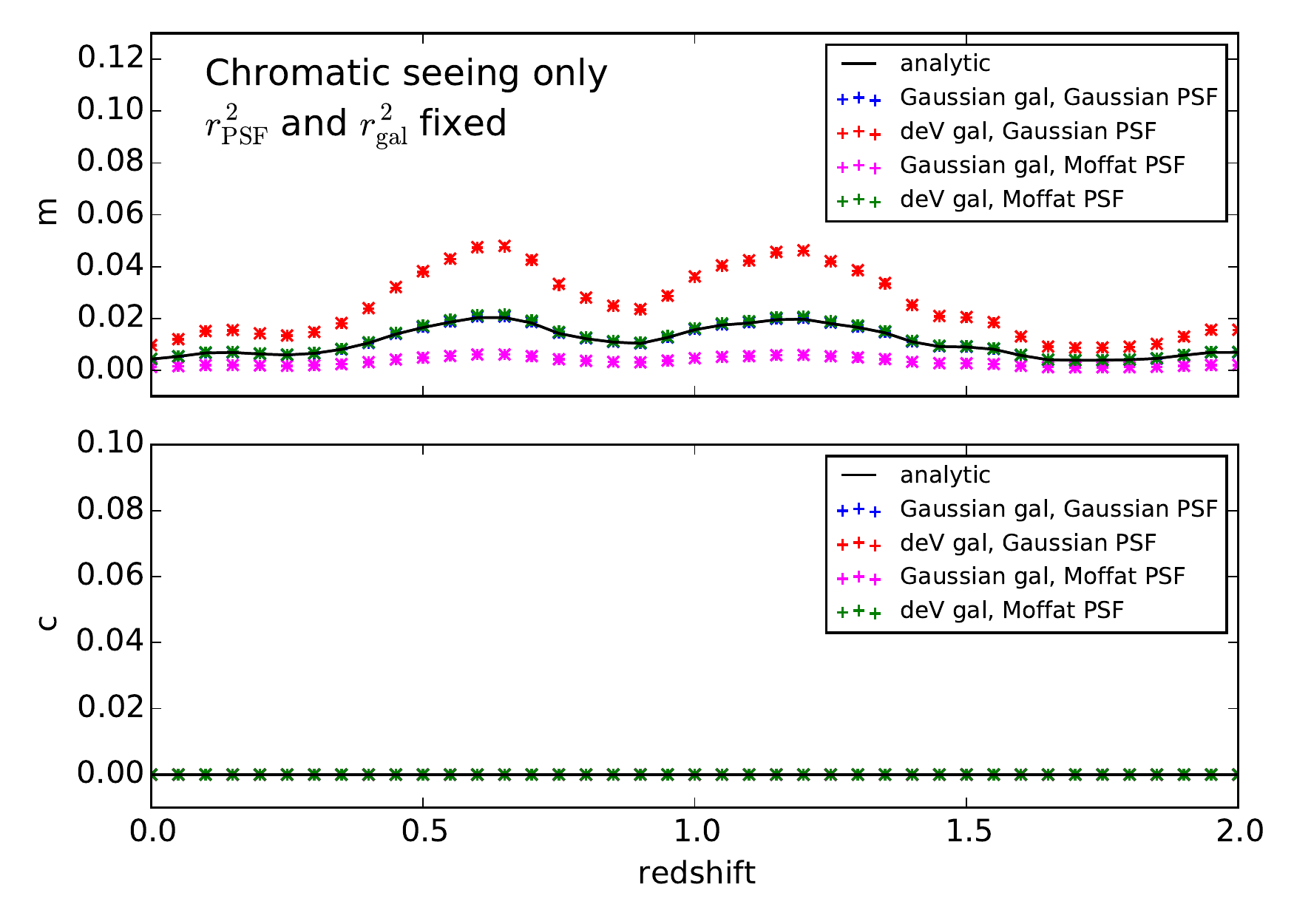}
  \end{center}
  \caption{
    \label{fig:r2_CS_modelfit}
    The shear calibration parameters due to chromatic seeing with no DCR.
    Details are the same as described in the caption for Figure~\ref{fig:r2_DCR_modelfit}.
    However, here the analytic prediction depends on both the galaxy and PSF second-moment squared radii, and is otherwise independent of the profiles of the PSF and galaxy.
  }
\end{figure}

\begin{figure}
  \begin{center}
    \epsscale{1.2}
    \plotone{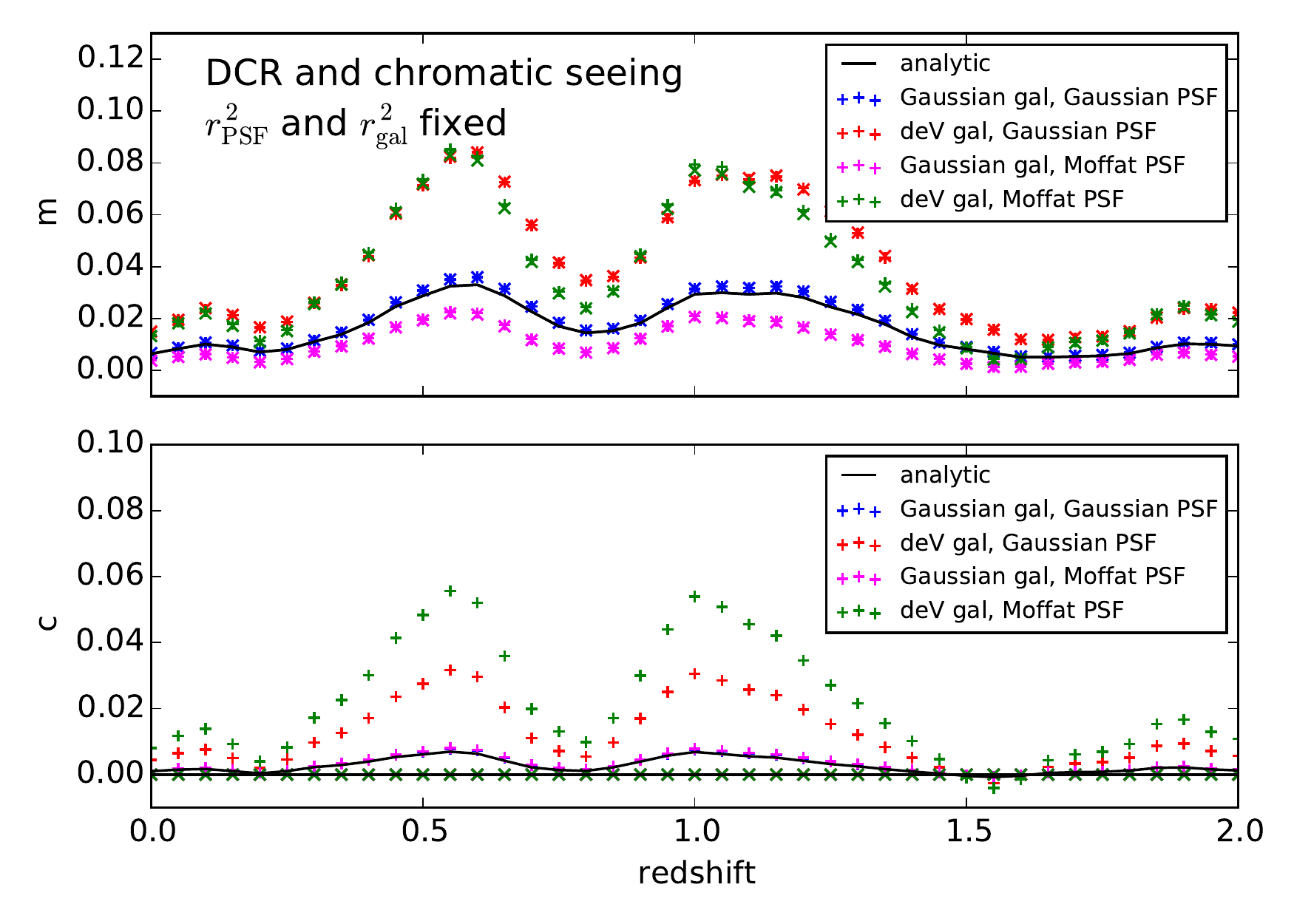}
  \end{center}
  \caption{
    \label{fig:r2_both_modelfit}
    The shear calibration parameters $m$ and $c$ due to both differential chromatic refraction and chromatic seeing.
    Details are the same as described in the caption for Figure~\ref{fig:r2_DCR_modelfit}.
    However, here the analytic prediction depends on both the galaxy and PSF second-moment squared radii, and is otherwise independent of the profiles of the PSF and galaxy.
  }
\end{figure}

We begin by investigating the dependence of chromatic effects on the shapes of the PSF and galaxy profile while holding the {\em second-moment squared radii fixed}.
Since the analytic formulae depend only on the second-moment squared radii, the same analytic predictions apply to all the investigated PSF and galaxy profiles.
In Figures~\ref{fig:r2_DCR_modelfit}-\ref{fig:r2_both_modelfit}, we compare analytic predictions for the shear calibration parameters $m$ and $c$, from Equations~\ref{eqn:m1m2_DCR}-\ref{eqn:c2_DCR} for DCR and Equations~\ref{eqn:m1m2_seeing}-\ref{eqn:c2_seeing} for chromatic seeing, to results from the ring test.

In Figure~\ref{fig:r2_DCR_modelfit}, the predicted shear calibration parameters are plotted as a function of redshift with only the physics of differential chromatic refraction included.
Here we see that the ring-test predictions for $m$ and $c$ for the Gaussian PSF and Gaussian galaxy profile agree quite well with the analytic predictions.
This might be expected, since the additivity of second moments under convolution (Equation~\ref{eqn:second_moments_add}), in addition to holding for unweighted second moments, also holds for the the very specific case where the PSF and galaxy profiles are Gaussian and the individual weight functions are identical to the individual profiles.
Our procedure in step 9 of finding the best least-squares fit to the Gaussian image is mathematically identical to measuring the moments with a matched Gaussian weight function \citep{Bernstein+Jarvis2002}.
Changing the PSF from a Gaussian to a Moffat shape also has little effect, which might be expected since Equations~\ref{eqn:m1m2_DCR}, \ref{eqn:c1_DCR}, and \ref{eqn:c2_DCR} have no dependence on the second moments of the PSF.
On the other hand, changing the shape of the galaxy from a Gaussian to a de~Vaucouleurs profile produces biases that are 2 to 6 times larger than predicted analytically, depending on the PSF profile.

Figure \ref{fig:r2_CS_modelfit} shows the predicted shear calibration parameters with only the physics of chromatic seeing included.
Again the ring-test predictions for the Gaussian PSF and Gaussian galaxy profile agree quite well with the analytic predictions.
Changing the PSF from a Gaussian to a Moffat profile reduces the predicted bias compared to its analytic value by a factor of $\sim$ 3, while changing the galaxy profile from Gaussian to de~Vaucouleurs increases the bias by a factor of $\sim$ 3.
Changing both the PSF and the galaxy profile results in a bias very similar to that predicted analytically.
As expected for chromatic seeing with a circular PSF, the predicted values of the additive biases ($c_i$) are zero, for both the analytic calculations and in the ring tests.

Figure \ref{fig:r2_both_modelfit} includes the physics of both differential chromatic refraction and chromatic seeing.
The ring-test predictions for a Gaussian PSF and Gaussian galaxy profile again show agreement with the analytic predictions, but for all other shapes considered, the predictions do not agree.

\begin{figure}
  \begin{center}
    \epsscale{1.2}
    \plotone{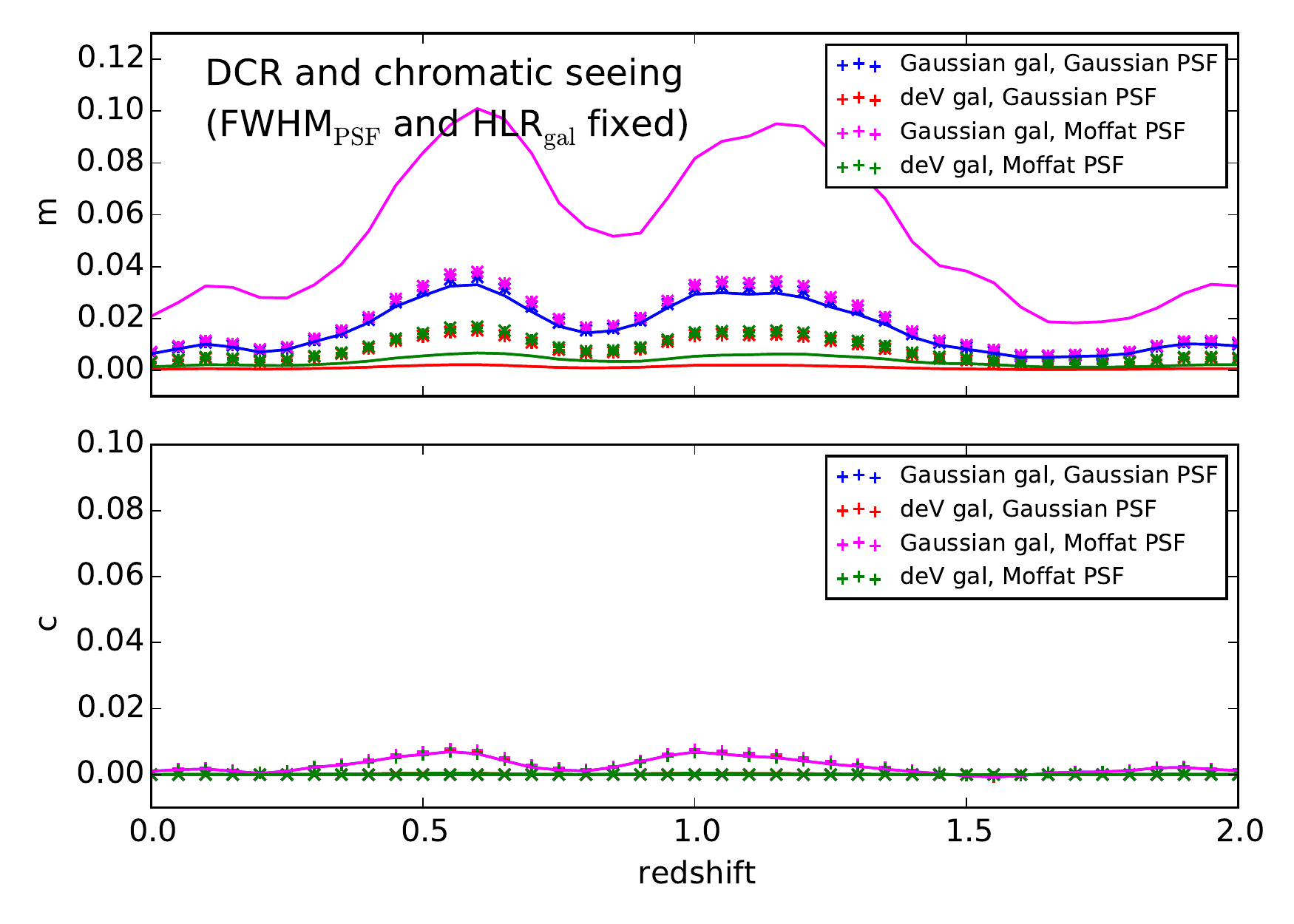}
  \end{center}
  \caption{
    \label{fig:FWHMHLR_both_modelfit}
    The shear calibration parameters $m$ and $c$ due to both differential chromatic refraction and chromatic seeing.
    Details are the same as described in the caption for Figure~\ref{fig:r2_DCR_modelfit}, except that here we hold fixed the PSF FWHM and the galaxy half-light radius, while investigating different assumed profiles.
    The solid curves represent the analytic predictions, which now depend on the choice of PSF and galaxy profile, and are colored to match the corresponding symbols for the ring-test predictions.
  }
\end{figure}

For the interpretation of Figures \ref{fig:r2_DCR_modelfit}, \ref{fig:r2_CS_modelfit}, and \ref{fig:r2_both_modelfit}, it is important to keep in mind that our choice to hold the galaxy and PSF second-moment squared radii fixed, as opposed to some other measure of size, was motivated primarily by the explicit appearance of this measure in the analytic formulae for chromatic biases.
This choice leads to some oddities in other measures of PSF or galaxy profile size.
For example, a $\beta=3.0$ Moffat PSF with the same second-moment squared radius as a Gaussian PSF will have a FWHM only 60\% as large as that of the Gaussian.
Similarly, a de~Vaucouleurs profile with the same second-moment squared radius as a Gaussian profile has a half-light radius four times smaller than that of the Gaussian.
Since PSF sizes are usually measured in terms of FWHM, and galaxy catalogs frequently report sizes in terms of half-light radii, we have plotted the analytic and ring-test--derived shear calibration parameters (for the case including both DCR and chromatic seeing) in Figure~\ref{fig:FWHMHLR_both_modelfit}, holding fixed the PSF FWHM and galaxy half-light radius, rather than the PSF and galaxy second-moment squared radii.
The main effect of this alternate, measurement-motivated choice of size measure is that the de~Vaucouleurs galaxies are now much larger for the same fiducial Gaussian profile galaxies than when the second-moment squared radii are matched, and the chromatic biases for these cases are therefore significantly reduced.
The change in PSF-size measure leads to greater consistency between the predicted biases for the two choices of PSF profile (Gaussian and Moffat).

In Table~\ref{table:profile_dependence}, we list the values of the shear calibration parameters due to both DCR and chromatic seeing, derived from analytic formulae and from the ring test for a G5V stellar SED and an Sa galactic SED, for a galaxy redshift of 0.6, which corresponds to a locally maximum bias in Figures~\ref{fig:r2_DCR_modelfit} - \ref{fig:FWHMHLR_both_modelfit}.
We list the biases for different combinations of PSF and galaxy-profile shapes, holding fixed the second-moment squared radii of the PSF and galaxy  (top group of three rows), and the PSF FWHM and galaxy half-light radius (bottom group of three rows).

\begin{deluxetable*}{lllllllllllll}
  \tablecaption{
    \label{table:profile_dependence}
    Analytic and ring-test predictions for shear bias parameters $m$ and $c$ due to differential chromatic refraction and chromatic seeing, for different PSF and galaxy profile shapes and sizes.
  }
  \tablehead{
    \multicolumn{2}{c}{Profile} &
    \multicolumn{2}{c}{PSF size} &
    \multicolumn{3}{c}{Galaxy size} &
    \multicolumn{2}{c}{Analytic prediction} &
    \multicolumn{2}{c}{Ring-test prediction} \\
    \colhead{PSF\phantom{pppppp}} &
    \colhead{Galaxy} &
    \colhead{$\mathrm{FWHM}_\mathrm{PSF}$} &
    \colhead{$\sqrt{r^2_\mathrm{PSF}}$\phantom{ppp}} &
    \colhead{$\mathrm{HLR}_\mathrm{gal}$} &
    \colhead{$\sqrt{r^2_\mathrm{gal}}$} &
    \colhead{$\mathrm{FWHM}_{\otimes}$\phantom{p}} &
    \colhead{$m_\mathrm{a}$} &
    \colhead{$c_{1,\mathrm{a}}$\phantom{ppppp}} &
    \colhead{$m_\mathrm{r}$} &
    \colhead{$c_{1,\mathrm{r}}$}
    }
  \startdata
% PSFtype    galtype    PSFFWHM       PSFr2         galHLR         galr2         convFWHM       m_a      c_a      m_r      c_r
  \multicolumn{8}{l}{\phantom{{\Large I}0000}{\it Fiducial PSF and galaxy profiles.}} \\
  Gaussian & Gaussian & $0.7\arcsec$  & $0.42\arcsec$ & $0.25\arcsec$ & $0.30\arcsec$ & $0.89\arcsec$ & 0.0330 & 0.0063 & 0.0359 & 0.0073\phantom{\LARGE I} \\
  \multicolumn{8}{l}{\phantom{{\Large I}0000}{\it Change profiles while fixing $r^2_\mathrm{PSF}$ and $r^2_\mathrm{gal}$.}} \\
  Gaussian & deV      & $0.7\arcsec$  & $0.42\arcsec$ & $0.06\arcsec$ & $0.30\arcsec$ & $0.76\arcsec$ & 0.0330 & 0.0063 & 0.0837 & 0.0296\phantom{\LARGE I} \\
  Moffat   & Gaussian & $0.43\arcsec$ & $0.42\arcsec$ & $0.25\arcsec$ & $0.30\arcsec$ & $0.74\arcsec$ & 0.0330 & 0.0063 & 0.0248 & 0.0074\\
  Moffat   & deV      & $0.43\arcsec$ & $0.42\arcsec$ & $0.06\arcsec$ & $0.30\arcsec$ & $0.51\arcsec$ & 0.0330 & 0.0063 & 0.0789 & 0.0431\\
  \multicolumn{8}{l}{\phantom{{\Large I}0000}{\it Change profiles while fixing $\mathrm{FWHM}_\mathrm{PSF}$ and $\mathrm{FWHM}_\otimes$.}} \\
  Gaussian & deV      & $0.7\arcsec$  & $0.42\arcsec$ & $0.46\arcsec$ & $2.16\arcsec$ & $0.89\arcsec$ & 0.0006 & 0.0001 & 0.0081 & 0.0042\phantom{\LARGE I} \\
  Moffat   & Gaussian & $0.7\arcsec$  & $0.69\arcsec$ & $0.21\arcsec$ & $0.25\arcsec$ & $0.89\arcsec$ & 0.0956 & 0.0090 & 0.0551 & 0.0109\\
  Moffat   & deV      & $0.7\arcsec$  & $0.69\arcsec$ & $0.30\arcsec$ & $1.38\arcsec$ & $0.89\arcsec$ & 0.0032 & 0.0003 & 0.0136 & 0.0054\\
  \multicolumn{8}{l}{\phantom{{\Large I}0000}{\it Change profiles while fixing $\mathrm{FWHM}_\mathrm{PSF}$ and $\mathrm{HLR}_\mathrm{gal}$.}} \\
  Gaussian & deV      & $0.7\arcsec$  & $0.42\arcsec$ & $0.25\arcsec$ & $1.16\arcsec$ & $0.84\arcsec$ & 0.0022 & 0.0004 & 0.0149 & 0.0071\phantom{\LARGE I} \\
  Moffat   & Gaussian & $0.7\arcsec$  & $0.69\arcsec$ & $0.25\arcsec$ & $0.30\arcsec$ & $0.95\arcsec$ & 0.0669 & 0.0063 & 0.0379 & 0.0074\\
  Moffat   & deV      & $0.7\arcsec$  & $0.69\arcsec$ & $0.25\arcsec$ & $1.16\arcsec$ & $0.87\arcsec$ & 0.0045 & 0.0004 & 0.0163 & 0.0064
  \enddata
  \tablecomments{
    These results assume a G5V stellar SED and an Sa galactic SED at redshift 0.6, which corresponds to a locally maximum bias in Figures~\ref{fig:r2_DCR_modelfit} - \ref{fig:FWHMHLR_both_modelfit}.
    The zenith angle for DCR calculations is 45 degrees.
    The ring test results assume that the intrinsic ellipticity of the galaxy is $\epsilon = 0.3$ (the results do not appreciably depend on the intrinsic ellipticity).
    The shorthand `deV' indicates a de~Vaucouleurs profile (Sérsic index $n=4$).
    $\mathrm{HLR}_\mathrm{gal}$ indicates the half-light radius of the galaxy profile.
    $\mathrm{FWHM}_{\otimes}$ indicates the FWHM of the convolution of the PSF and the galaxy profile.
    For the multiplicative and additive shear calibration parameters $m$ and $c$, a subscript `a' indicates an analytic result, and a subscript `r' indicates a result derived from a ring test.
    Note that $m_1$ and $m_2$ are precisely equal in the analytic formulae, and differ by less than $0.0006$ for the ring-test results, so we simply report the average value.
    Similarly, since we assume that the monochromatic PSF is circular, and that the `1' direction is along the zenith, all $c_2$ values are 0.
  }
\end{deluxetable*}

We also include in Table~\ref{table:profile_dependence} the shear calibration parameters for the case when the PSF FWHM and the FWHM of the convolution of the PSF and the galaxy profile ($\mathrm{FWHM}_{\otimes}$) are held fixed (middle group of three rows), which corresponds to the approach adopted by \citet{Voigt++12} and others.
We find that this size description leads to just as complicated a dependence of the shear calibration parameters on PSF and galaxy profiles as the size descriptions mentioned above.

Focusing on the last three rows in Table~\ref{table:profile_dependence}, we see that the analytic predictions can be significantly different than the ring test results, and that both can lead to significant systematic biases compared to the expected DES and LSST cosmic shear statistical uncertainties -- $|\langle m\rangle|_\mathrm{max}$ in Table~\ref{table:surveys}.
(Also note that the vertical-axis range of Figures \ref{fig:r2_DCR_modelfit}-\ref{fig:FWHMHLR_both_modelfit} are more than an order of magnitude larger than the requirements given in Table \ref{table:surveys}.)
The dependence on galaxy profile is also a concern, since it implies that a correction derived from only stellar and galactic photometry, as suggested by \citet{Cypriano++10} and PB12, will not be sufficient for mitigating chromatic biases if applied at the catalog level.

\subsection{Ring-Test Case Study}\label{sec:casestudy}

\begin{figure*}
  \begin{center}
    \epsscale{1.0}
    \plotone{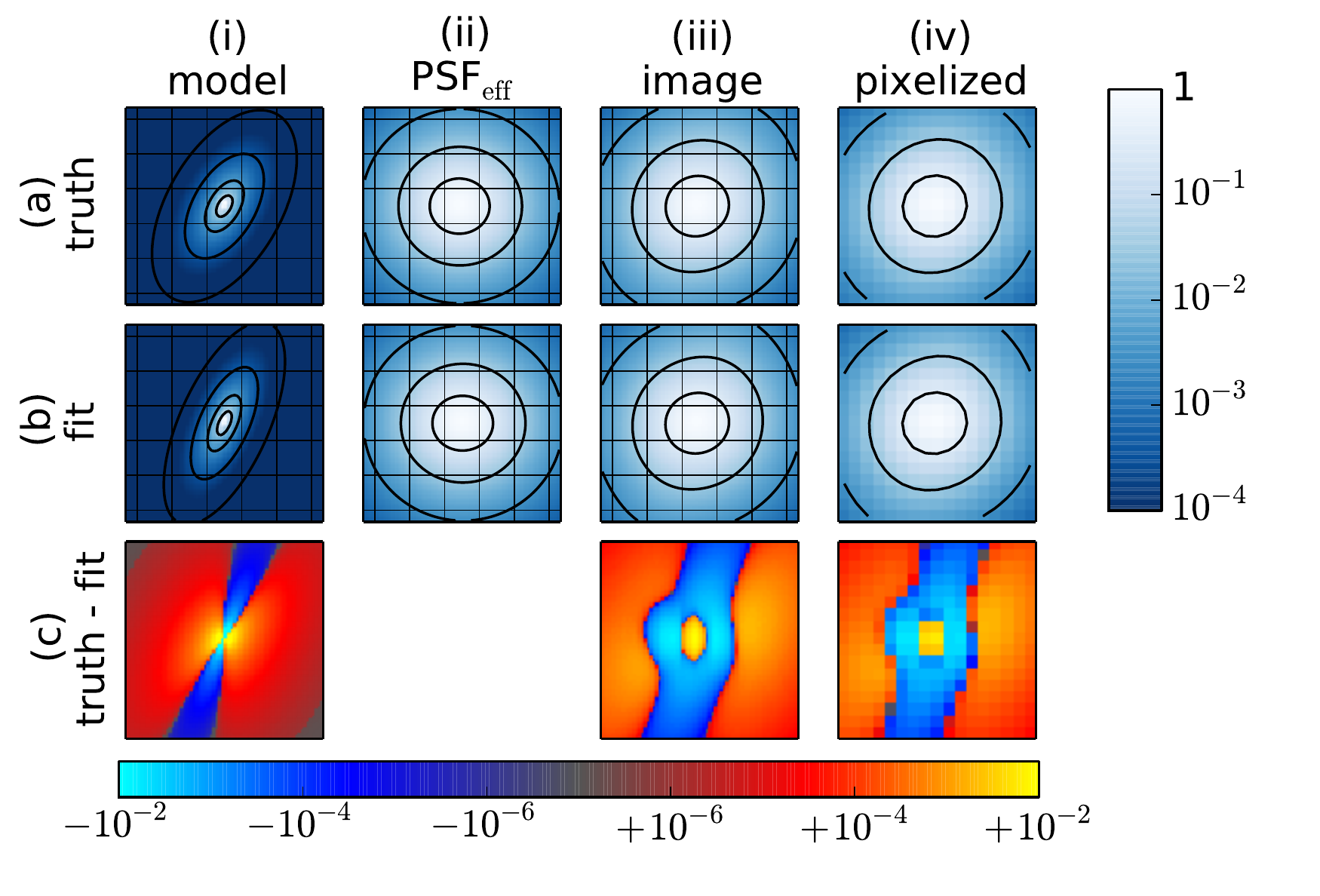}
  \end{center}
  \caption[fig2]{
    \label{fig:diagnostic}
    Illustration of the simulation and fitting portion of the ring-test procedure (steps 8 and 9 in Section~\ref{sec:ringtest}).
    The columns from left to right represent
    (i) the high-resolution model galaxy image before PSF convolution,
    (ii) the high-resolution effective PSF,
    (iii) the galaxy image convolved with the PSF, and
    (iv) the pixelation of (iii).
    The rows from top to bottom represent
    (a) the target galaxy model, galactic PSF, convolution, and pixelation,
    (b) the best-fit galaxy model, stellar PSF, convolution, and pixelation, and
    (c) the residuals ($\mathrm{truth} - \mathrm{best\ fit}$).
    Note the separate symmetric logarithmic axis for the color scale in the third row, with red corresponding to positive residuals and blue to negative residuals.
    The fit in step 9  minimizes the squared residuals, summed over pixels, for the image in the bottom right corner.
    The difference between the target and best-fit galaxy models (bottom left corner) illustrates the model bias induced by using the stellar PSF in the fit.
    In this figure, the galaxy profile is de~Vaucouleurs (Sérsic index $n = 4.0$) both in the target image and fixed during the fit, the monochromatic PSF is a $\beta=3.0$ Moffat profile with the effective galactic PSF derived from an Sa spectrum at redshift 0.6, and the effective stellar PSF derived from the spectrum of a G5V star.
    Both the physics of chromatic seeing and differential chromatic refraction are included.
    The zenith angle is set to 60 degrees, and galaxy size set to $r^2_\mathrm{gal}=(0.3\arcsec)^2$ to help visualize the effects.
}
\end{figure*}

To investigate the ring-test results in more detail, we conduct a case study focusing on the redshift 0.6 galaxy with an Sa SED, for which the predicted shear calibration biases are largest, for both the analytic and ring-test predictions.
We construct a series of diagnostic plots (Figure~\ref{fig:diagnostic}) that illustrate the simulation and fit results described in steps 8 and 9 of the ring test.
The effective stellar and galactic PSFs include both the physics of chromatic seeing and differential chromatic refraction.
These diagnostic figures are generated for a zenith angle of 60 degrees and a somewhat smaller galaxy with $r_\mathrm{gal}^2 = (0.3\arcsec)^2$ for better visualization, though the qualitative results are the same for smaller zenith angles or larger galaxies.

In Figure~\ref{fig:diagnostic}, the first row illustrates (from left to right) the ingredients in step 8: the generated galaxy profile, the effective PSF for an Sa galactic SED, the convolution of the galactic PSF and the `true' galaxy profile, and the pixelated image.
The second row illustrates (from left to right) the results of step 9: the pre-convolution best-fit galaxy model, the effective PSF for a G5V stellar SED, the convolution of the stellar PSF and the best-fit galaxy model, and the best fit to the pixelated target image.
The rightmost panel in the third row shows the residual between the target image and the best fit image, and is an indication of the quality of the fit in step 9 of the ring test.
If the fit were perfect, the pixel values in the residual image would be uniformly 0.

The residual image is not uniformly 0 due to `model bias' \citep{Melchior++09, Voigt+Bridle10, Bernstein10}.
The fitting step can be viewed as an attempt to deconvolve the target image by the stellar PSF under the assumption that the functional form of the deconvolved image is known.
The best-fit model (second row, first column) is the result of this attempt to deconvolve the image.
The ellipticity-parameter estimators can then simply be calculated from the parameters describing the best-fit model.
In practice, however, the deconvolution of the target image by the stellar effective PSF cannot be represented by the functional form of the fit (and hence is not shown in any of the panels in Figure~\ref{fig:diagnostic}).
The true deconvolution (as opposed to the model-fit approximate deconvolution) may not even have elliptical isophotes, precluding a solution that simply includes more degrees of freedom in the description of the radial profile in the functional form.
Degrees of freedom can, of course, be added {\it azimuthally} (see, for example, \citet{Ferrari++04} or \citet{Peng++10}) to potentially obtain a perfect deconvolution via model fitting, but then the ellipticity is no longer identically -- or uniquely -- a model parameter.

The degree to which the ring-test predictions for shear calibration biases are influenced by these model-fitting limitations depends on how well the deconvolution is able to perform.
Larger residuals between the best-fit and true profiles (lower left panel in Figure~\ref{fig:diagnostic}) indicate more model bias.
Several shape-measurement algorithms have been proposed specifically to address model bias~\citep{Bernstein10, Melchior++11}.
These algorithms, however, aim to mitigate bias due to using an incorrect model for the galaxy profile, whereas the model bias we see is due to using an incorrect PSF.

\section{PSF-level correction} %7
\label{sec:psf_corr}
Earlier studies of chromatic PSF effects, particularly those pertaining to the space mission {\it Euclid}, propose to calibrate the average multiplicative shear bias as a function of observables such as galaxy color and redshift \citep{Voigt++12, Semboloni++13}.
This calibration can then be applied a posteriori to the measured ellipticity of each galaxy.
However, the dependence of atmospheric chromatic effects on the PSF and galaxy profiles (coming from model-fitting bias) implies that the atmospheric chromatic effects studied here cannot be realistically corrected at the `catalog level'.
We propose instead to correct individual stellar and galactic effective PSFs before measuring galaxy ellipticities.
This strategy is more efficient for ground-based experiments since the atmospheric PSF can vary significantly from one exposure to the next.
An a posteriori catalog-level calibration similar to that proposed for {\it Euclid} would depend in a complicated way on all of the PSFs of the individual exposures.
Correcting the individual PSFs also has the advantage of being independent of all non-photometric properties of galaxies -- in particular these corrections are independent of galaxy shapes.

\subsection{Method}
\label{sec:psf_corr_method}

Our approach is to apply small perturbations to the PSF model derived from stars to yield a galactic PSF model that is applicable to the deconvolution of an individual galaxy image.
The perturbations we apply depend on both the physics of the chromatic effect involved and the photometry of the stars and galaxies under consideration.
The PSF is typically only sparsely sampled by suitable stars across an image and must be interpolated to the positions of galaxies in weak lensing analyses.
We study whether chromatic effects can be corrected during the PSF-interpolation stage through the following ordered steps.

\begin{enumerate}

\item % 1
For each stellar image that will be used to measure the PSF, estimate the effective PSF at the location of the star. The details of this estimate are nontrivial and are beyond the scope of this paper.

\item % 2
Correct for differences in differential chromatic refraction between the measured stellar effective PSF and a fiducial effective PSF with specified SED by {\em deconvolving}\footnote{This may need to be a {\em convolution} by a Gaussian with second moment
$V^\mathrm{fid} - V^\mathrm{*}$
if $V^\mathrm{*} < V^\mathrm{fid}$.
We suggest that the fiducial PSF have a monochromatic SED, in which case $V^\mathrm{fid}=0$, to avoid this complication.} the stellar effective PSF in the zenith direction by a Gaussian with second moment $V^\mathrm{*} - V^\mathrm{fid}$
and first moment $\bar{R}^\mathrm{*} - \bar{R}^\mathrm{fid}$, which can be estimated from the stellar photometry via a machine-learning algorithm as shown in Section~\ref{sec:machine_learning}.

\item % 3
Correct for differences in chromatic seeing between the measured stellar effective PSF and the fiducial effective PSF by {\em scaling} the coordinate axes of the PSF model from step 2 by $r^2_\mathrm{PSF,fid}/r^2_\mathrm{PSF,*}$, which can also be estimated from photometry via a machine-learning algorithm as shown in Section~\ref{sec:machine_learning}.
Note that in at least some analytic PSF models, such as Gauss-Laguerre decomposition, this step and the previous step (and also steps 5 and 6 below) can be implemented analytically.

\item % 4
Interpolate the fiducial monochromatic PSF model samples to the positions of the galaxies.
This step is also nontrivial and is beyond the scope of this paper.

\item % 5
For each galaxy, reverse step 3 by scaling the PSF coordinate axes by $r^2_\mathrm{PSF,g}/r^2_\mathrm{PSF,fid}$, which can also be estimated from photometry via a machine-learning algorithm.

\item % 6
For each galaxy, reverse step 2 by convolving (or deconvolving) the PSF in the zenith direction by a Gaussian with second moment $V^\mathrm{gal} - V^\mathrm{fid}$ and first moment $\bar{R}^\mathrm{gal} - \bar{R}^\mathrm{fid}$, which can also be estimated from photometry via a machine-learning algorithm.

\end{enumerate}

An exact correction of differential chromatic refraction in steps 2 and 6 would amount to a deconvolution or convolution in the zenith-direction by the DCR kernel given in Equation \ref{eqn:convker} and shown in Figure \ref{fig:photon_landings}.
Since the detailed DCR kernel depends on the detailed SED of the particular star or galaxy over the wavelength range of the filter, which is generally unknown, we instead approximate this kernel by a Gaussian with our best estimate of the correct second moment.
The second moment in the zenith direction for the resulting DCR-corrected PSF will be almost correct (up to the precision of the machine-learning correction, and not yet accounting for chromatic seeing).
Similarly, steps 3 and 5 yield our best estimate (up to the precision of the machine-learning correction) of the second-moment correction for chromatic seeing.
While we could attempt to correct moments higher than second by also learning these from photometry, we will see that this appears to be unnecessary.

\subsection{Calibrating the Corrections}
\label{sec:psf_corr_calibrate}

In Section \ref{sec:ETR}, we found that the Extra Trees machine-learning algorithm was able to predict the mean DCR bias parameters (which introduce multiplicative biases to shear measurements) to beyond the precision required to keep systematic uncertainties below statistical uncertainties (see Figure \ref{fig:corrected_bias_panel}).
This statement implicitly assumes, however, that the distribution of SEDs in the real universe matches the distribution of SEDs in \textsc{CatSim}.
One can check this assumption by acquiring a sample of unbiased SEDs covering the wavelength range spanned by the $r$- and $i$-band filters.
These SEDs can also be used to calibrate the machine-learning output by applying a correction equal to the difference between the mean predicted chromatic bias output by the \textsc{CatSim}-trained machine-learning algorithm applied to real photometry and the mean chromatic bias obtained by integrating observed SEDs.
From the residual variance of DCR parameters derived using the \textsc{CatSim} SED distribution, we estimate that with just a few spectra per redshift bin (assuming $\sim$~10 redshift bins) one should be able to \emph{measure} the mean residual DCR biases to a precision well below the DCR requirements and remove them from the analysis.
For chromatic seeing, the residual square-root-variance is about twice the requirement, indicating that we need at least four unbiased SEDs per redshift bin to calibrate the systematic errors to the level of the statistical errors, and probably at least a few dozen SEDs per bin to calibrate well beyond this level of precision.

For additive errors, we remind the reader that the variance requirements we derive in this paper, which assume that additive errors are maximally correlated, are sufficient but not necessary to keep additive systematic uncertainties below statistical uncertainties.
Future studies incorporating realistic clustering of galaxies with different SEDs will be able to set requirements directly on the additive bias correlation function.
Since the variance estimates we make in this paper are less than an order-of-magnitude above the sufficient variance requirement for LSST, we expect that such studies will reveal that additive systematics from atmospheric chromatic effects will not strongly affect cosmological constraints.
We therefore leave strategies to further mitigate residual additive bias to future work.

\subsection{Results}
\label{sec:psf_corr_results}

\begin{figure}
  \begin{center}
    \epsscale{1.2}
    \plotone{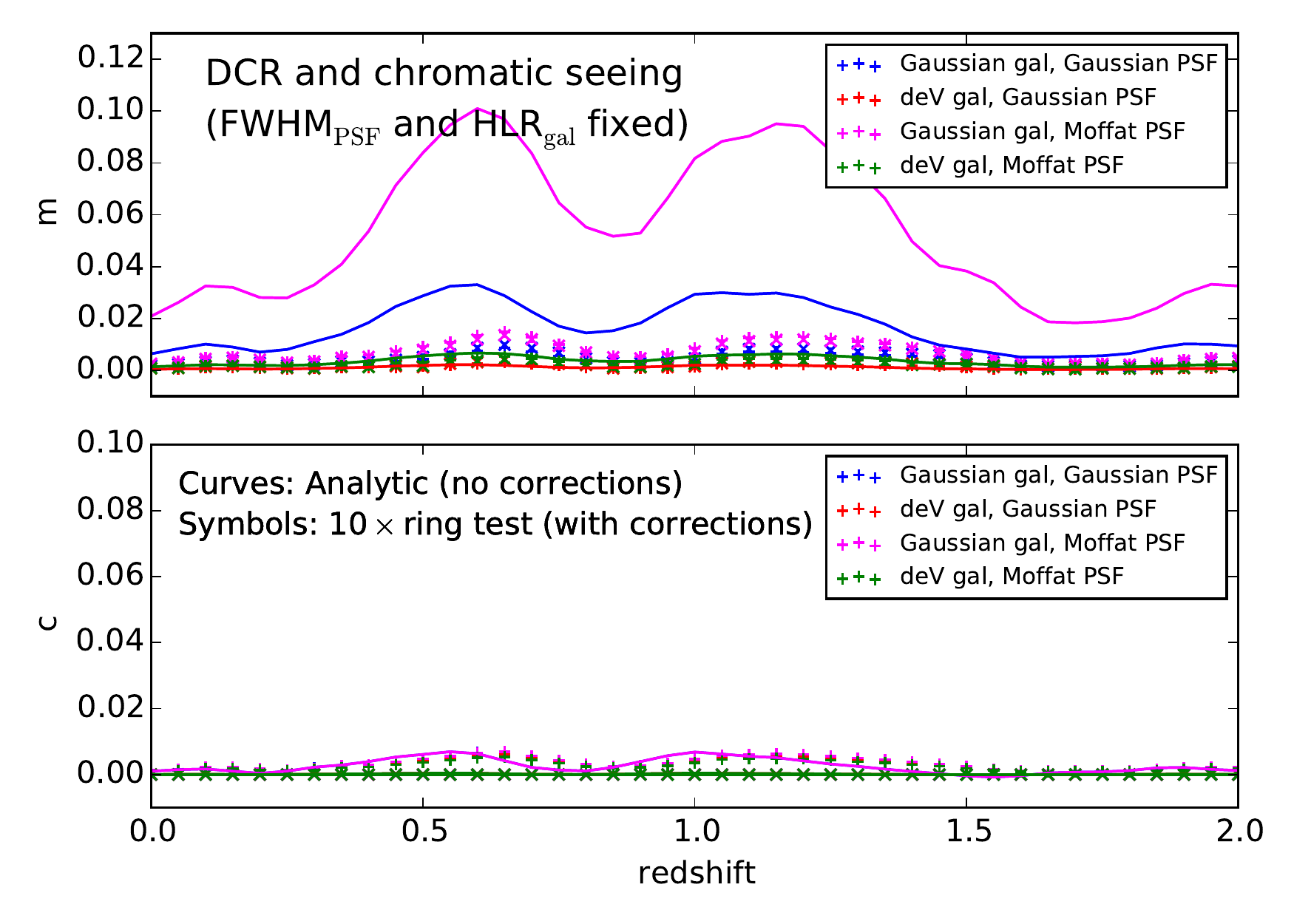}
  \end{center}
  \caption{
    \label{fig:FWHMHLR_both_perturb}
    The shear calibration parameters $m$ and $c$ due to both differential chromatic refraction and chromatic seeing.
    The colored curves correspond to 1$\times$ the analytic predictions for the {\em uncorrected} biases.
    The symbols correspond to 10$\times$ the ring-test predictions {\em after} the perturbative corrections to the PSFs described in steps 1 through 6 in Section~\ref{sec:psf_corr} are applied.
    Details are the same as described in the caption for Figure~\ref{fig:r2_DCR_modelfit}, except that we hold fixed the PSF FWHM and galaxy half-light radius while investigating different assumed profiles (as in Figure~\ref{fig:FWHMHLR_both_modelfit}).
    The vertical scale is the same as that in Figure~\ref{fig:FWHMHLR_both_modelfit}.
    The largest residual chromatic biases after the PSF-level corrections are $|m| \sim 0.0015$ and $|c| \sim 0.0007$; these can be compared to the requirements given in  Table~\ref{table:surveys} and described in the Note below the table.
  }
\end{figure}

We implemented the correction scheme described in Section \ref{sec:psf_corr_method} (without calibrating to real data) and tested it by performing a ring test using the corrected PSFs in step 9 of the algorithm presented in Section \ref{sec:ringtest}.
We investigate the same set of PSF and galaxy profiles, holding the PSF FWHM and galaxy half-light radius fixed, as those for which the shear calibration parameters are shown in Figure~\ref{fig:FWHMHLR_both_modelfit} and in the last three rows in Table~\ref{table:profile_dependence}.
To isolate chromatic effects and our ability to correct for model-fitting bias, we do not include effects related to PSF estimation or interpolation, but simply correct the exact stellar effective PSF using the exact values of $\Delta V$ and $\Delta r^2_\mathrm{PSF}/r^2_\mathrm{PSF}$.
The results are shown in Figure~\ref{fig:FWHMHLR_both_perturb}, which is plotted with the same scale as
Figure~\ref{fig:FWHMHLR_both_modelfit} for comparison, and in Table~\ref{table:PSF_corr} for a G5V stellar SED and an Sa galactic SED at redshift 0.6.
Note that, in Figure~\ref{fig:FWHMHLR_both_perturb}, the values of $m$ and $c$ for the symbols, which correspond to the ring-test predictions {\em after} the perturbative corrections to the PSF are applied, are multiplied by a factor of 10 so that their distribution is visible relative to the analytic predictions for the {\em uncorrected} biases (colored curves).
The largest residual chromatic biases after the PSF-level corrections are $|m| \sim 0.0015$ and $|c| \sim 0.0007$.
These biases can be compared to the requirements given in  Table~\ref{table:surveys} and described in the Note below the table.

We remind the reader that the combination of stellar and galactic SEDs used in this experiment is particularly challenging in terms of the chromatic biases produced, and that the simulated zenith angle of 45 degrees is at the 80th percentile of the expected zenith angle distribution for LSST.
We therefore anticipate that the residual chromatic biases for the majority of LSST shear measurements will be significantly less than those measured here.

\begin{deluxetable*}{llllll}
  \tablecaption{
    \label{table:PSF_corr}
    Ring-test predictions for shear bias parameters $m$ and $c$ before and after PSF correction.
  }
  \tablehead{
    \multicolumn{2}{c}{Profile}\phantom{pppppp} &
    \multicolumn{2}{l}{Ring-test prediction} &
    \multicolumn{2}{l}{After PSF correction} \\
    \colhead{Galaxy} &
    \colhead{PSF\phantom{ppppppppp}} &
    \colhead{$m_\mathrm{r}$} &
    \colhead{$c_{1,\mathrm{r}}$\phantom{ppppppppp}} &
    \colhead{$m_\mathrm{c}$} &
    \colhead{$c_{1,\mathrm{c}}\phantom{ppppppppp}$}
    }
  \startdata
% galtype    PSFtype    m_r      c_r      m_c      c_c
  Gaussian & Gaussian & 0.0359 & 0.0073 & 0.0007 & 0.0007\\
  deV      & Gaussian & 0.0149 & 0.0071 & 0.0003 & 0.0006\\
  Gaussian & Moffat   & 0.0379 & 0.0074 & 0.0013 & 0.0007\\
  deV      & Moffat   & 0.0163 & 0.0064 & 0.0005 & 0.0005
  \enddata
  \tablecomments{
    These results assume a G5V stellar SED and an Sa galactic SED at redshift 0.6, and include both DCR (zenith angle of 45 degrees) and chromatic seeing.
    The ring test results assume that the intrinsic ellipticity of the galaxy is $\epsilon = 0.3$.
    For the shear calibration parameters $m$ and $c$, a subscript `r' indicates a result derived from a ring test based on an uncorrected stellar PSF, and a subscript `c' indicates a result derived from a ring test based on the perturbative corrections to the PSFs described in steps 1 through 6 in Section~\ref{sec:psf_corr}.
    As the profile of the PSF and galaxy profile are varied, the PSF FWHM and galaxy half-light radius are held fixed at $0.7$ arcsec and $0.25$ arcsec, respectively -- i.e., the same as the values listed in the top row and bottom three rows of Table~\ref{table:profile_dependence}.
    Since $m_1$ and $m_2$ differ by less than $0.0006$ for the ring test results, we simply report their average.
    Similarly, since we assume that the monochromatic PSF is circular, and that the `1' direction is along the zenith, all $c_2$ values are zero.
    The corrected value of $m$ can be compared to the value $|\langle m\rangle|_\mathrm{max} = 0.003$ from Table~\ref{table:surveys} as the value at which the combined uncertainties from all systematic effects will equal the expected statistical uncertainty of the LSST weak-lensing survey.
    The corrected value of $c$ can be compared to the value $\sqrt{\mathrm{Var}(c)_\mathrm{max}} = 3 \times 10^{-4}$ which is a conservative upper limit on the point at which all systematic effects will equal the expected statistical uncertainty of the LSST weak-lensing survey.
  }
\end{deluxetable*}

\section{Discussion of limitations, other chromatic effects, and broader impacts} %8
\label{sec:discussion}
We have described the impact of wavelength-dependent PSFs on galaxy shape measurements for two chromatic effects, both due to the atmosphere: differential chromatic refraction and chromatic seeing.
In this section, we discuss uncertainties or limitations in our study that could affect the predictions, other potential (non-atmospheric) contributions to chromatic biases, and broader impacts of wavelength-dependent PSFs for the LSST survey and image analysis.

\subsection{Uncertainties and Limitations in Predictions of Atmospheric Chromatic Bias}

{\em Differential chromatic refraction:}
The physics of DCR is quite well understood. However, the impact of DCR on galaxy shape measurements in a particular survey depends in a nontrivial way on the distribution of zenith and parallactic angles at which any patch of the sky is observed over an entire survey;
see Section~\ref{sec:DCRfirstmoment}.
Therefore, chromatic bias due to DCR is impacted by survey strategy.

Additionally, our results are sensitive to the typical source galaxy second-moment squared radius $r_{\rm gal}^2$.
As can be seen from Equations\,\ref{eqn:m1m2_DCR} and \ref{eqn:c1_DCR}, for given requirements on the shear bias parameters $m$ and $c$, the requirements on the mean and variance of shifts in second central moment $\Delta V$ due to DCR are proportional to $r_{\rm gal}^2$ and $r_{\rm gal}^4$, respectively, as are the requirements on the mean and variance of shifts in $(\Delta \bar{R}_{45})^2$.
The estimates for typical $r_{\rm gal}^2$ in Table~\ref{table:sizes} are sensitive to precisely how the galaxy size is extracted from an image and to selection criteria for galaxy properties such as flux and signal-to-noise ratio.

{\em Chromatic seeing:}
In this paper, we have assumed that i) the PSF wavelength dependence due to seeing is a linear isotropic dilation, and ii) the wavelength dependence of the size of the dilation follows a power law relation:
$\theta \propto \lambda^p$.
With the value of $p=-0.2$ that we have used, these assumptions correspond to pure Kolmogorov turbulence in the infinite exposure limit.
However, the validity of both i) and ii) for the 15-second exposures that LSST will use appears to be an open question.

A handful of studies have investigated the value of the exponent of the expected power law relation.
The only direct measurement of which we are aware comes from observations of the solar limb taken over a wide range of wavelengths (0.55\,$\micron$ to 10\,$\micron$) and with bad seeing (FWHM of $5$\,arcseconds at 0.55\,$\micron$)~\citep{Boyd78}.
For these conditions, the authors of the study report results consistent with $p = -0.22 \pm 0.04$.

The value of $p$ can also be indirectly estimated by observing the power spectrum of refractive index fluctuations or temperature fluctuations.
With this method, inferred values of $p$ vary from the Kolmogorov prediction ($-0.2$) to significantly smaller values \citep[$\sim -0.4$,][]{Linfield++01}.

The weak-lensing multiplicative bias $m$ scales roughly linearly with $p$.
For LSST, we have seen that with no correction applied for chromatic biases, the multiplicative bias for $p=-0.2$ is about a factor of $10$ larger than survey requirements.
Therefore, to correct this bias for LSST (e.g., with machine-learning algorithms based on simulations), we need to know $p$ to a precision of
$\Delta p = 0.02$ or better to ensure that the remaining uncertainty does not dominate the statistical uncertainties.

We are not aware of any empirical studies that test the extent to which monochromatic PSFs at different wavelengths are related via a linear isotropic dilation over moderate exposure times.
Such a study might be accomplished using a seeing-limited integral field spectrograph.

Finally, our results are sensitive to the typical ratio of source galaxy to PSF second-moment squared radius $r_{\rm gal}^2 / r_{\rm PSF}^2$.
As can be seen in Equations\,\ref{eqn:m1m2_seeing}, \ref{eqn:c1_seeing}, and \ref{eqn:c2_seeing}, for given requirements on the shear bias parameters $m$ and $c$, the requirements on the mean and variance of differences in star-galaxy PSF sizes $\Delta r^2_\mathrm{PSF} / r^2_\mathrm{PSF}$ due to chromatic seeing are proportional to $r_{\rm gal}^2 / r_{\rm PSF}^2$ and $r_{\rm gal}^4 / r_{\rm PSF}^4$, respectively.

{\em Color gradients:}
In Section~\ref{sec:analytics}, Equation~\ref{eqn:no_color_gradients}, we made the assumption that the spatial and wavelength dependence of galaxy surface brightness profiles are separable -- i.e., there are no color gradients.
An accurate prediction of biases in cosmic shear estimators due to color gradients requires a high-fidelity model of color gradients in the sample of galaxies used to measure weak lensing.
We leave the treatment of color gradients for ground-based surveys to future work.

{\em System throughput:}
Our proposed PSF corrections require a detailed understanding of the system throughput including the atmosphere, optics, filters, and sensors.
Section 2.6 of the LSST Science Book \citep{LSSTSB} details the planned strategy for calibrating and monitoring the LSST system throughput, including the use of monochromatic dome flats \citep{Stubbs+Tonry06} and a 1.2-m auxiliary calibration telescope used to measure the contemporaneous atmospheric transmission \citep{Stubbs++07}.

\subsection{Other Chromatic Effects That Contribute to the PSF}
\label{sec:other_chroma}

Other contributions to wavelength-dependence of the PSF, such as chromatic effects in the optics or the sensors, can also introduce biases in shape measurements.
The requirements on how precisely these chromatic effects must be known can depend on details such as any variation of the effect across the image plane.
Assessing these potential contributions to cosmic-shear bias will benefit from, and may require, precise measurements of the wavelength dependence of the sensor response, and high-fidelity simulations of chromatic effects in the optical system and the sensor response.
We also point out that not all chromatic effects will be readily broken down into wavelength-dependent shifts and dilations, and thus the correction strategy in Section \ref{sec:psf_corr} may need to be generalized for some effects.

Here is a (possibly incomplete) list of additional chromatic PSF effects:
\begin{itemize}
\item Chromatic aberration from refractive optics, which can lead to an effect similar to DCR, except that stretching of the PSF will exhibit radial symmetry in the focal plane.
\item Diffraction.
\item Optical aberrations.
\item Optical distortion (which is not part of the PSF, but can lead to epoch misregistration similar to DCR centroid shifts).
\item Charge diffusion in the CCD -- longer wavelength photons convert farther into the silicon and the resulting electrons will diffuse less.
\item Lateral electric fields near the edge of the CCD coupled with the wavelength dependence of photon conversion depth.
\item Charge repulsion in the CCD coupled with the wavelength dependence of photon conversion depth.
\item Spatially non-uniform filter throughput.
\end{itemize}

\subsection{Other Impacts of Wavelength-Dependent PSFs}

{\em Image processing:} The fact that differential chromatic refraction depends on the zenith angle of the observation and affects both the first and second moments of the PSF in an SED-dependent manner has implications for the processing of images in the LSST survey.
In order to determine accurate astrometry in a single image and achieve precise registration between images with procedures similar to those outlined  in this paper (i.e., PSF corrections determined from machine-learning algorithms based on six-band photometry), photometric information must be available for every object. Each object has a unique PSF correction that depends not only on the position of the object in the image, but also on the object's SED and the zenith and parallactic angles of the observation.
This has implications for the design of the analysis pipeline, especially in a `multifit' environment when many exposures are being fit simultaneously.

{\em Simulations:} The corrections based on machine learning described in Sections~\ref{sec:machine_learning} and \ref{sec:model_fitting} require training catalogs containing photometry and predicted chromatic biases.
To generate such a catalog, we need high-fidelity simulations of chromatic effects in the atmosphere, optics, and detectors, for a population of stars that are representative in their SEDs, and galaxies in their SEDs, sizes, redshifts, and color gradients, for different observing conditions.
One may also hope to calibrate chromatic effects directly from the on-sky data, though this may be challenging considering the small size of the effects.

{\em Transient detection:} Wavelength-dependent PSFs, particularly due to DCR, will impact the image subtraction that will be used in the (nearly) real time LSST pipeline designed to rapidly detect interesting transient events.
Each new image will be compared to a deep template.
Differences in chromatic biases between the new image and the template can lead to false detections of transient events in the difference image.
Requirements and solutions to these challenges are being actively explored by others (A. Becker, private communication).

\section{Open-source software tools} %9
\label{sec:software}
Documented software and scripts developed for the studies described in this paper are available in the open-source LSST DESC github repository \url{https://github.com/DarkEnergyScienceCollaboration/chroma}.
As part of this work, chromatic effects relevant to these studies have been implemented in the open-source \textsc{GalSim} galaxy simulation package \citep[\url{https://github.com/GalSim-developers/GalSim},][]{Rowe++14}.
These tools can be used to replicate and extend the studies presented here -- for example, to explore different star and galaxy catalogs, different survey strategies, or other filters.  
The chromatic effects implemented in the \textsc{GalSim} package include surface brightness profiles with inseparable dependence on position and wavelength, allowing the study of chromatic biases for different models of galaxy color gradients.

\section{Summary and outlook} %10
\label{sec:conclusions}
Because of chromatic effects in the atmosphere, using the point spread function measured with stars to determine the shape of a galaxy that has a different spectral energy distribution than the stars leads to biases in galaxy shape measurements that, if uncorrected, degrade the constraining power of cosmic shear estimators by more than the predicted statistical precision of future ground-based astronomical surveys.
We summarize here the dominant effects and how they might be corrected.

\begin{enumerate}

\item Chromatic seeing introduces multiplicative shear biases that are equivalent to or exceed the statistical precision of the full DES and LSST surveys, respectively, in both $r$- and $i$-band.
(See bottom panel in Figure~\ref{fig:bias_panel} in Section~\ref{sec:catalog}.)

\item Chromatic biases due to differential chromatic refraction are significantly larger in $r$-band than $i$-band.
  At a zenith angle of 45 degrees, the chromatic shifts in second central moments introduce multiplicative systematic uncertainties equivalent to the statistical precision of the full LSST survey in $r$-band.
(See top panel in Figure~\ref{fig:bias_panel} in Section~\ref{sec:catalog}.)

\item The variances of chromatic biases derived from the LSST \textsc{CatSim} SED catalog greatly exceed the sufficient but not necessary requirements we derive for DES and LSST.
  (See middle panel in Figure~\ref{fig:bias_panel} in Section~\ref{sec:catalog}.)
  However, future work incorporating realistic clustering of galaxies with different SEDs may show that requirements on additive bias correlation functions are met.

\item Machine-learning techniques applied to six-band LSST photometry can be used to predict chromatic PSF biases.
  (See Figure~\ref{fig:corrected_bias_panel} in Section~\ref{sec:machine_learning}.)
  However, these predicted biases are not easily propagated into galaxy shape biases, which we find also depend on the galaxy and PSF profiles -- so-called `model bias' -- in addition to galactic and stellar SEDs.

\item Atmospheric chromatic biases can be predicted and reduced with minimal model bias by applying an ordered set of perturbative PSF-level corrections based on machine-learning techniques applied to six-band photometry.
(See Figure~\ref{fig:FWHMHLR_both_perturb} and Table~\ref{table:PSF_corr} in Section~\ref{sec:psf_corr}.)

\end{enumerate}

These studies show that even small wavelength dependencies in the PSF can introduce biases in galaxy shape measurements.
Hence, achieving the ultimate precision for weak lensing from current and future imaging surveys will require a detailed understanding of the wavelength dependence of the PSF from all sources, including the atmosphere, the optics, and the CCD sensors.
Minimizing and correcting chromatic biases impacts the design of the survey and the data analysis pipeline, and requires high-fidelity simulations.
We conclude that, if uncorrected, chromatic biases in shear measurements in LSST can be larger than the expected statistical uncertainty.
Catalog-level corrections are unlikely to work due to model-fitting biases.
PSF-level corrections will be necessary and, for the atmospheric chromatic effects studied here, sufficient to achieve the LSST requirements for galaxy shape measurements.

\acknowledgments
This work was supported by National Science Foundation grant PHY-0969487 and PHY-1404070.
We thank the developers of the \textsc{GalSim} galaxy simulation package and Michael Jarvis, Rachel Mandelbaum, and Barney Rowe, in particular, for advice and feedback on implementing chromatic effects in \textsc{GalSim}.
We thank those contributing to the LSST simulation efforts -- in particular Simon Krughoff for use of the simulated catalog of stars and galaxies (\textsc{CatSim}), and those who produced the simulated observing strategy for a possible ten-year LSST survey (\textsc{OpSim}).
We thank David Kirkby for his helpful notes on computing signal-to-noise ratios.
We thank Gary Bernstein, James Bosch, {\v Z}eljko Ivezi{\'c}, Michael Jarvis, David Kirkby, and Tony Tyson for their helpful feedback on this paper.

\appendix
\section{Photometric uncertainties} %A
\label{sec:magerr}
In Section \ref{sec:machine_learning}, we investigate the susceptibility to photometric uncertainties of the machine-learning algorithm used to predict chromatic biases.
Here we present our algorithm for estimating the photometric uncertainties for LSST.

For a given filter, the total detected signal rate $s$ for each star or galaxy can be written in terms of its AB magnitude $m$ and the filter zeropoint $s_0$,
\begin{equation}
  \label{eqn:signal}
  s=s_0 10^{-0.4 \left(m-m_0\right)},
\end{equation}
where $m_0$ is a reference magnitude, the units of $s$ and $s_0$ are electrons per second, and we are ignoring any airmass dependence.
This signal occurs on top of a relatively large sky background $B$ with units of magnitudes per square arcsecond.
The sky rate in electrons per second per pixel is
\begin{equation}
  \label{eqn:sky}
  S=s_0 10^{-0.4\left(B-B_0\right)}\cdot \Delta^2_\mathrm{pix},
\end{equation}
where $B_0 = m_0 \cdot \mathrm{arcsec}^{-2}$ and $\Delta^2_\mathrm{pix}$ is the solid angle subtended by one pixel.
For lensing source galaxies and PSF stars, the dominant source of noise is Poisson fluctuations of the sky background.
Using an optimal weight function for photometry (i.e., one which has the same profile as the object being measured) yields a signal-to-noise ratio of
\begin{equation}
  \label{eqn:SNR}
  \nu = \sqrt{\frac{t_\mathrm{exp}}{S} \sum_{\mathrm{pixel}\, p} \mathbf{D}^2_p},
\end{equation}
where $t_\mathrm{exp}$ is the total exposure time and $\mathbf{D}$ indicates the number of electrons per pixel in the noise-free image of the galaxy.
The magnitude uncertainty is related to the signal-to-noise by
\begin{equation}
  \label{eqn:magerr}
  \sigma(m) = \frac{2.5}{\log 10 \cdot \nu}.
\end{equation}

\begin{figure}
  \begin{center}
    \epsscale{1.0}
    \plotone{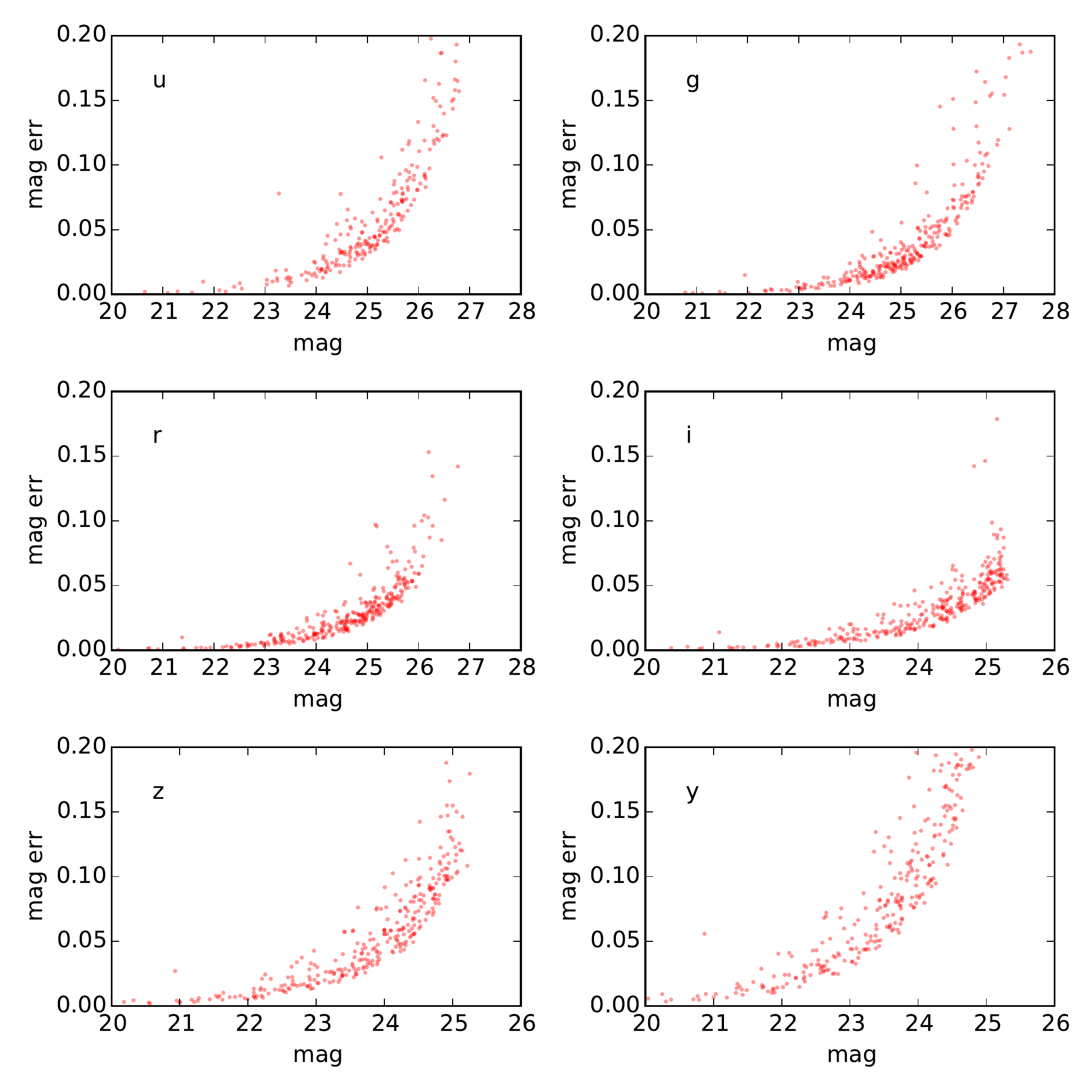}
  \end{center}
  \caption{
    \label{fig:magerr}
    Fiducial 10-year LSST magnitude uncertainties for $i < 25.3$ \textsc{CatSim} galaxies in $u$-, $g$-, $r$-, $i$-, $z$-, and $y$-bands.
  }
\end{figure}

Because the magnitude uncertainties depend not only on each object's magnitude but also on the surface brightness profile, we compute the uncertainty for each individual object.
We use \textsc{GalSim} to draw a PSF-convolved noise-free image of each galaxy or star in the \textsc{CatSim} catalog and evaluate equation \ref{eqn:SNR} to obtain the signal-to-noise ratio and subsequently the magnitude uncertainty.
The sky brightness, zeropoint, total number of visits for each LSST filter, and the resulting total exposure times used for the calculation are listed in Table \ref{table:snr_params}.
Figure \ref{fig:magerr} shows the resulting relation between magnitude and uncertainty for each filter in the fiducial 10-year LSST survey.

\begin{deluxetable*}{lllll}
  \tablecaption{
    \label{table:snr_params}
    Sky brightness and exposure depths for LSST in different filters.
  }
  \tablehead{
    \colhead{Filter} &
    \colhead{$B (\mathrm{mag}\cdot\mathrm{arcsec}^{-2})$} &
    \colhead{$s_0 (e^-/{\mathrm s})$} &
    \colhead{$N_\mathrm{visit}$} &
    \colhead{$t_\mathrm{exp} (\mathrm{s})$}
    }
  \startdata
  % Filter           B      s_0      N_exp      t_exp
  u & 22.8 & 24.3 & 56  & 1680 \\
  g & 22.2 & 70.6 & 80  & 2400 \\
  r & 21.3 & 55.9 & 180 & 5400 \\
  i & 20.3 & 41.5 & 180 & 5400 \\
  z & 19.1 & 28.6 & 164 & 4920 \\
  y & 18.1 & 15.0 & 164 & 4920
  \enddata
  \tablecomments{
    Sky brightness estimates $B$ are from \citet{Gee++07}.
    Zeropoints $s_0$ are estimated by \citet{Kirkby++inprep} and assume reference magnitude of $m_0 = 24$.
    The number of visits to each patch of sky in each filter $N_\mathrm{visit}$ is from \citet{LSSTSB}, and the total exposure time $t_\mathrm{exp}$ is estimated from $N_\mathrm{visit}$, assuming two 15-second exposures per visit.
  }
\end{deluxetable*}

\section{Additive Requirements} %A
\label{sec:addreq}
Additive systematic shear biases $c$ influence cosmological inferences through their correlation functions or power spectra.
In this paper, however, we estimate only the mean and variance of $c$ as a function of redshift.
In this Appendix we connect the correlation function of $c$ to the mean and variance of $c$.

The cross-correlation function of the complex field $c$ between tomographic redshift bins $i$ and $j$ is
\begin{equation}
  \label{eqn:corrfunc}
  \xi^{c,ij}_+(\theta) = \Re[\langle c_i c_j^* \rangle _\theta],
\end{equation}
where the subscript $\theta$ indicates that the average is to be taken over all pairs separated by an angle $\theta$ on the sky, and the real-part operator is necessary because we do not generally expect systematic effects to be parity-invariant. (The real-part operator is usually omitted in the expression for the shear correlation function, since shear is expected to be parity-invariant and thus its correlation function is real automatically.)
One can also view Eq. \ref{eqn:corrfunc} as the real part of the sum of the covariance of pairs $c_i$ and $c_j$ and the product of their means:
\begin{equation}
  \label{eqn:xi_split}
  \xi^{c,ij}_+(\theta) = \Re[\Covar(c_i, c_j)_\theta + \langle c_i \rangle \langle c_j^* \rangle],
\end{equation}
where the covariance of two complex-valued random variables is
\begin{equation}
  \Covar(a, b) = \langle (a - \langle a \rangle) (b - \langle b \rangle)^* \rangle.
\end{equation}
The weak lensing literature is not always clear about whether or not the term $\Re[\langle c_i \rangle \langle c_j^* \rangle]$ in Eq. \ref{eqn:xi_split} should be included in the definition of the correlation function.
Regardless, this term is independent of $\theta$, and hence only affects the additive systematic power spectrum at $\ell=0$.
To bound the power spectrum at $\ell > 0$, one can derive a bound on the covariance part of Eq. \ref{eqn:xi_split} given in terms of the variance of each component of $c$ in each tomographic bin:
\begin{equation}
    \left|\Re[\Covar(c_i, c_j)]\right| \le \sqrt{\Var(\Re(c_i))\Var(\Re(c_j))} + \sqrt{\Var(\Im(c_i))\Var(\Im(c_j))}.
\end{equation}

In Sections \ref{sec:catalog} and \ref{sec:machine_learning}, we estimate the variances of $\Delta V$ and $\Delta r^2_\mathrm{PSF}/r^2_\mathrm{PSF}$ under the simplifying assumption that these quantities are real -- i.e., we ignore their directional variation.
(The same is not true for $\Delta (\bar{R}_{45})^2$ -- here we estimate both components in Figure \ref{fig:misregistration_bias}.)
To correct for this, we assume that with directional variation included, the imaginary component may be as large as the real component, and hence the bound in terms of just the real component is
\begin{equation}
  \left|\Re[\Covar(c_i, c_j)]\right| \le 2 \sqrt{\Var(\Re(c_i))\Var(\Re(c_j))}.
\end{equation}
Note that this approximation is likely pessimistic, since including the directional variation also reduces the variance of the real component.
This factor of 2 has been incorporated into the $\Delta V$ and $\Delta r^2_\mathrm{PSF}/r^2_\mathrm{PSF}$ requirements in Table \ref{table:chromatic_requirements}, and Figures \ref{fig:bias_panel} and \ref{fig:corrected_bias_panel}.
Finally, since the geometric mean of two positive quantities is less than or equal to the larger of the two quantities, we can write
\begin{equation}
  \left|\Re[\Covar(c_i, c_j)]\right| \le 2\,\mathrm{max}(\Var(\Re(c_i))) \equiv C_\mathrm{max}.
\end{equation}

The additive bias power spectrum is related to the additive bias correlation function via
\begin{equation}
  C^{ij}_c(\ell) = 2\pi\int_0^\infty{\theta\,\xi^{c,ij}_+(\theta) J_0(\ell \theta) \dif \theta}.
\end{equation}
\citet{Amara+Refregier08} set a requirement on the integrated power spectrum of additive shear systematics,
\begin{equation}
  \sigma^2_\mathrm{sys} = \frac{1}{2 \pi}\int_{\ell_\mathrm{min}}^{\ell_\mathrm{max}}{|C^\mathrm{sys}(\ell)|(\ell+1) \dif \ell},
\end{equation}
such that the systematic uncertainty in the dark energy equation of state parameter $w$ is less than or equal to the statistical uncertainty for weak lensing experiments.
For their chosen integration limits of $\ell_\mathrm{min} = 10$ and $\ell_\mathrm{max} = 20000$, they estimate the requirement $\sigma^2_\mathrm{sys} \lesssim 3 \times 10^{-7} (1 \times 10^{-7})$ for DES (LSST).
Similarly, \citet{Massey++13} extend the \citet{Amara+Refregier08} analysis by removing the ambiguity of integration limits and tomographic binning by defining
\begin{equation}
  \bar{\mathcal{A}} = \frac{\sum\nolimits_{ij}\frac{1}{2\pi}\int_{\ell_\mathrm{min}}^{\ell_\mathrm{max}}{|C_{ij}^\mathrm{sys}(\ell)|\ell \dif \ell}}{\sum\nolimits_{ij}\frac{1}{2\pi}\int_{\ell_\mathrm{min}}^{\ell_\mathrm{max}}{\ell \dif \ell}}.
\end{equation}
Their estimated requirement for the {\it Euclid} weak lensing survey, which has similar statistical properties to LSST, is $\bar{\mathcal{A}} < 1.8\times10^{-12}$.
Note that both $\sigma^2_\mathrm{sys}$ and $\bar{\mathcal{A}}$ are independent of $\langle c_i \rangle$ since this only affects the power spectrum at $\ell=0$.

To relate $C_\mathrm{max}$ to $\sigma^2_\mathrm{sys}$ or $\bar{\mathcal{A}}$, we consider a correlation function (setting the term that is constant in $\theta$ to zero) that is maximum up to some angular size $\theta_\mathrm{max}$ and 0 at larger scales:
\[
\xi^{c,ij}_+(\theta) =
\begin{cases}
  C_\mathrm{max}, & \theta \le \theta_\mathrm{max}, \\
  0, & \theta > \theta_\mathrm{max}.
\end{cases}
\]
While this correlation function does not rigorously maximize $\sigma^2_\mathrm{sys}$ or $\bar{\mathcal{A}}$ under the constraint of a given $C_\mathrm{max}$, it is quite pessimistic compared to realistic systematic correlation functions so long as $\theta_\mathrm{max}$ is at least a few degrees.
The power spectrum of this pessimistic correlation function can be written analytically:
\begin{equation}
  C^{ij}_c(\ell) = C_\mathrm{max} \frac{\theta_\mathrm{max} J_1(\ell \theta_\mathrm{max})}{2 \pi \ell}.
\end{equation}
This power spectrum can then be integrated to yield values of $\sigma^2_\mathrm{sys}$ and $\bar{\mathcal{A}}$ in terms of $C_\mathrm{max}$, which are listed in Table \ref{table:add} for several choices of $\ell_\mathrm{min}$, $\ell_\mathrm{max}$, and $\theta_\mathrm{max}$.
For this paper, we propagate requirements on additive atmospheric chromatic biases using the value $C_\mathrm{max} = 1.8 \times 10^{-7}$, derived from the estimated requirement $\bar{\mathcal{A}} < 1.8 \times 10^{-12}$ \citep{Massey++13} and assuming $\ell_\mathrm{min}=10$, $\ell_\mathrm{max}=5000$, and $\theta_\mathrm{max}=5^{\circ}$.
The equivalent requirement for DES is $\bar{\mathcal{A}} < 6.0 \times 10^{-12}$.

\begin{deluxetable*}{ccccccc}
  \tablecaption{
    \label{table:add}
    Additive systematic sufficiency requirements.
  }
  \tablehead{
    \colhead{$\ell_\mathrm{min}$} &
    \colhead{$\ell_\mathrm{max}$} &
    \colhead{$\theta_\mathrm{max}$} &
    \colhead{$\sigma^2_\mathrm{sys} / C_\mathrm{max}$} &
    \colhead{$C_\mathrm{max}$} &
    \colhead{$\bar{\mathcal{A}} / C_\mathrm{max}$} &
    \colhead{$C_\mathrm{max}$} \\
    \multicolumn{4}{l}{} &
    \colhead{AR08} &
    &
    \colhead{M13}
    }
  \startdata
  % lmin  lmax      thmax      sigma2   A
  $10$ & $5000$  & $5^\circ$  & $20$ & $5.0\times10^{-9}$ & $1.0\times10^{-5}$ & $1.8\times10^{-7}$ \\
  $10$ & $20000$ & $5^\circ$  & $42$ & $2.4\times10^{-9}$ & $1.3\times10^{-6}$ & $1.4\times10^{-6}$ \\
  $10$ & $5000$  & $10^\circ$ & $29$ & $3.4\times10^{-9}$ & $1.4\times10^{-5}$ & $1.3\times10^{-7}$
  \enddata
  \tablecomments{
    Columns $\sigma^2_\mathrm{sys} / C_\mathrm{max}$ and $\bar{\mathcal{A}} / C_\mathrm{max}$ report quite pessimistic values for the ratio of the additive systematic metrics described by \citet{Amara+Refregier08} (AR08) and \citet{Massey++13} (M13), respectively, to the maximum correlation function amplitude of additive shear calibration bias $c$, for various choices of $\ell_\mathrm{min}$, $\ell_\mathrm{max}$, and $\theta_\mathrm{max}$.
    Columns labeled $C_\mathrm{max}$ then use the estimated LSST requirements of $\sigma^2_\mathrm{sys} < 1\times10^{-7}$ and $\bar{\mathcal{A}} < 1.8 \times 10^{-12}$ to set conservative upper limits on the tolerable level of $C_\mathrm{max}$.
    Additive systematics with variance in $c$ smaller than these values will introduce systematic uncertainties in the dark energy equation of state parameter $w$ no larger than the statistical uncertainties.
    We note, however, that depending on the correlations of $c$, the actual requirements are likely to be much less strict.
  }
\end{deluxetable*}

\vskip0.5cm

\bibliographystyle{apj}
\bibliography{chroma}

\end{document}